# CORAL: A Modality-Invariant Framework for Robust Vital Sign Rate Estimation Using Correloform Analysis

Jacob Trueb [1], Amey Kasbe [1], and Aniruddh Srinivasan [1]

***Abstract*—Heart rate and respiration rate are crucial vital signs. We present CORAL to address challenges in automating continuous vital sign monitoring 1) in-hospital across broader populations (age, disease states), 2) at-home in telehealth and wellbeing applications (noise, placement), and 3) across data modalities (ECG, PPG, SCG, BioZ, etc.). We re-introduce Short-Time Autocorrelation Functions (STACFs) to introduce the correloform — a highly interpretable 2D signal transformation for tracking periodicity over time — and define CORAL as a generic analytical framework for robust rate estimation of quasi-periodic signals. We rigorously benchmark CORAL to show ubiquitous application in biosignals via capabilities arising by mathematical construction rather than domain-specific engineering, including rate estimation, resilient noise handling, automatic channel selection, and signal quality indication.**

**CORAL achieves excellent instantaneous noninvasive fetal HR agreement in FECGSYNDB (F1 $= 0.999$) and ADFECGDB (F1 $= 1.000$). With difficult NICU neonates, CORAL estimates RR from wearable BioZ at $r = 0.858$ against breath-interval RR from simultaneous wired Philips impedance. CORAL SCG HR on CEBS achieves $r = 0.990$ and in free-living activity $r = 0.970$ when compared against commercial ECG. Without QRS detection, CORAL's instantaneous HR agrees with Pan–Tompkins and NeuroKit2 detectors as closely as they agree with each other (MIMIC-IV ICU ECG $r = 0.87$ to each vs. $0.88$ between them). CORAL's HR standard deviation correlates strongly with interval-based HRV SDNN, even from mechanical SCG (CEBS $r = 0.733$) and across an ICU ECG cohort ($r = 0.768$).**

## I. INTRODUCTION

Vital signs provide essential summaries of a person's health. Heart rate (HR) and respiratory rate (RR) are particularly fundamental measures of cardiopulmonary function that challenge conventional estimation methods outside of routine clinical circumstances. Even in inpatient clinical settings, unhealthy, fetal, neonatal, and geriatric patients suffer from less accurate HR and RR estimation [1], [2]. At the same time, the rise of wearable sensing technology expands access to biosignals like electrocardiography (ECG), photoplethysmography (PPG), and seismocardiography (SCG) in telehealth and wellbeing applications. The quintessential wearable, the smartwatch, accounts for 568 million users in 2026 with 905 million users projected by 2031 [3], substantiating the growing worldwide integration of wearable technologies into everyday life. While current commercial wearable platforms provide signals acceptable for high-risk patients in hospital, they may exhibit less reliable HR and RR than currently deployed stationary clinical equipment [4]. Wearable platforms lack generalizable methods that reliably capture HR and RR with clinical-grade accuracy from distinct types of biosignals in widely varying patient types and settings.

Extrapolating automated HR estimation from ECG to PPG, SCG, and other sensing modalities is difficult even in healthy populations due to SNR differences and the physiological origin of the measured phenomena. Cardiopulmonary measures may exhibit additional static or dynamic morphology variation dependent on a variety of factors such as tissue composition, device placement, interface degradation, body orientation, and activity. HR estimates that rely on normative expectations further struggle in patients with pathophysiologies that significantly alter waveform morphology (atrial fibrillation, cardiac pacing, neonates with cardiopulmonary disease, etc.). Ongoing work pursues improving signal quality and accuracy to meet the unaddressed clinical need via advancements in machine learning such as with fetal HR, blood pressure, and arrhythmia [5], [6], [7]. Prior efforts also produced several signal quality indicators to identify suitable segments for processing [7], [8], [9]. RR estimation methods face additional challenges for reliable performance, owing to variety within gold-standard mechanisms — 1) mouth and nasal airflow, 2) chest and abdomen wall movement. Typical approaches require additional processing to isolate respiratory activity. Most automated methods to estimate HR and RR from biosignals are thus tailored to the expected input data type and output range of the population of interest.

Despite the wealth of differences between cardiopulmonary biosignals, they all share foundationally quasi-periodic outputs. The autocorrelation function quantifies the periodicity of a signal by calculating the correlation between a signal and its copy at shifted overlaps. Elevated periodicity in a signal manifests as increased correlations at non-zero offsets. With increases in continuous biosignal availability, autocorrelation re-emerged as a compelling tool for health-related applications. Recent work with autocorrelation-based algorithms demonstrated promise for SCG-derived HR in adults, post-operative recovery of homeostatic regulation in pediatric patients, and monitoring activity patterns in patients with Parkinson's Disease [10], [11], [12], [13]. To our knowledge, research efforts have yet to integrate the capabilities of autocorrelation into an algorithmic framework for biosignal and patient-naïve rate capture.

[1]Trueb, Kasbe, Srinivasan are with Querrey Simpson Institute for Bioelectronics, Querrey Simpson Institute for Translational Engineering for Advanced Medical Systems, Northwestern University, Technological Institute, A/B Wing, 2145 Sheridan Road Evanston, IL 60208, {jtrueb, amey.kasbe, ani}@northwestern.edu.

To address the challenge of generalizable HR and RR estimation, we developed a dynamic programming approach with minimal heuristic integration that calculates HR and RR from autocorrelation-derived biosignal periodicity. In this work, we introduce the correloform signal transformation and evaluate the analytical framework (CORAL) on clinical public benchmarking datasets and medical research device datasets. On the fetal benchmarks, CORAL achieves ML-range F1 without training data. We demonstrate that CORAL performs excellently for HR estimation on a variety of biosignals in patients ranging from *in utero* fetal patients to geriatric patients with valvular heart disease. We also showcase state-of-the-art capabilities with RR estimation from several biosignals recorded from patients spanning pre-term neonates to sleeping children. CORAL's adaptive signal quality index has broad utility; it varies systematically across cardiovascular diagnostic superclasses and automates removal of artifacts without requiring out-of-band information. Finally, we demonstrate CORAL's instantaneous rate estimation is sufficiently granular to correlate strongly with HRV in ICU bedside ECG, in diagnostic ECG strips, and SCG in controlled settings.

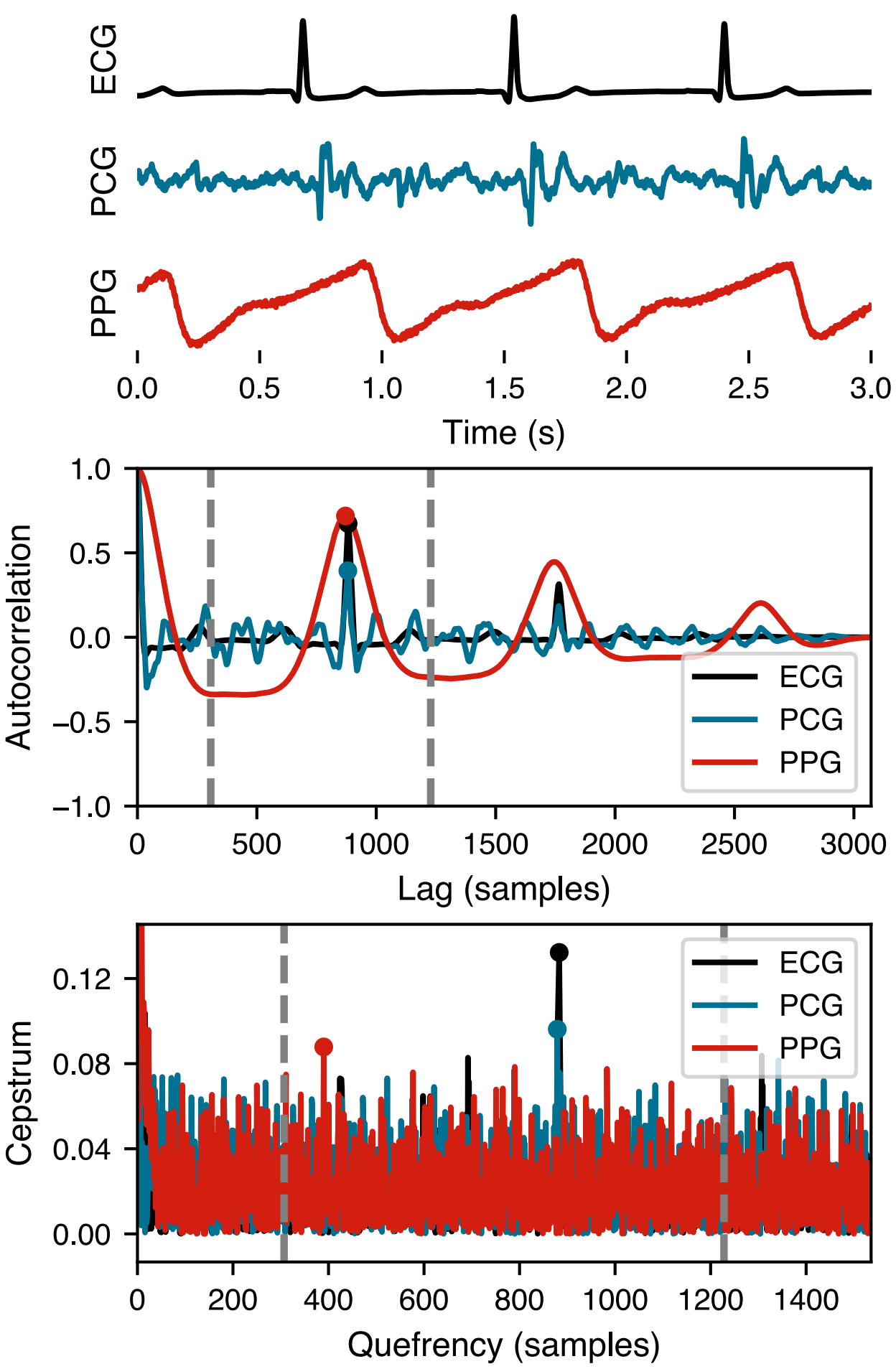

Fig. 1. Autocorrelation and Cepstrum of 1024 Hz ECG, PCG, and PPG signal with peaks within search bounds (50–200 bpm). Autocorrelation identifies the dominant RR peak interval at 70 bpm.

## II. LITERATURE REVIEW

### A. *Electrocardiography*

Introduced in 1902, ECG is a long-studied mechanism for tracking cardiac activity by measuring electrical depolarizations [14]. 10-second recordings of 12-lead ECG are standard-of-care, while single-lead ECG is becoming common in telehealth and continuous monitoring. Cardiologists manually identify established fiducial locations in ECG morphology to clinically inform decision making [15]. In ECG, methods for estimating HR rely heavily on QRS complex detection, operating on heuristics such as expected ECG contours and timing intervals during the sequence of depolarization on each heartbeat [16]. For example, the widely deployed Pan–Tompkins algorithm includes a QRS and T wave disambiguation step if detected peaks are within 360 ms of each other [16], [17], [18]. Respiration induces changes in ECG waveform morphology driven by Respiratory Sinus Arrhythmia (RSA), enabling estimation of RR. Analysis of ECG waveforms using 2D representations and deep learning for diagnosis and management is an active area of research [5].

### B. *Photoplethysmography*

PPG, first demonstrated in 1938, leverages the differential optical absorption spectra of oxygenated and deoxygenated hemoglobin to measure hemodynamic function [19]. The PPG waveform inversely varies with the presence of oxygenated hemoglobin, leading to a quasi-periodic sinusoidal peak morphology that varies with the mode of optical stimulation, patient skin pigmentation, and underlying cardiovascular processes. Well-established insights from PPG include HR, RR and $SpO_2$ [20]. Given that smartwatches rely on wrist PPG data for health insights, novel PPG processing methods offer promise for holistic improvement of population health. For example, [21] tried to send notifications when smartwatch PPG periodicity was poor and movement stopped to alert on cardiac arrest within 60 s. PPG-derived HR initially relied on temporal peak detection and interval aggregation for rate estimation [22]. PPG-derived RR leverages amplitude and frequency analysis of the individual waveform to estimate RR [9]. Motion artifacts are a notable challenge in PPG analysis, along with generally low precision due to the relatively gross measure of occlusion due to perfusion [22]. With the rapid integration of ML models into digital health, modern PPG processing relies on models trained on PPG and accelerometer data for motion-free data identification and subsequent metric estimation [23].

### C. Vibrocardiography

Vibrocardiography — including phonocardiography, ballistocardiography, SCG, gyrocardiography — is a family of vibration-based cardiac monitoring first demonstrated in the late 1800s [24], [25], [26], [27], [28], [29]. Made feasible by the rise of microelectromechanical systems (MEMS) and digital signal processing (DSP), SCG is a promising mechanism for monitoring cardiopulmonary rates in naturalistic and clinical environments. When placed on the torso, MEMS accelerometers capture the high frequency vibrations caused by heart valves opening and closing (20–150 Hz) and low frequency chest wall deflections associated with breathing (0.2–1.5 Hz). DSP enables separation of these frequency bands and enables HR and RR estimation. Wearable SCG devices developed by BioIntelliSense and Sibel Health underwent successful commercial translation in recent years [30], [31], [32], [33]. SCG offers comparative advantages to ECG and PPG for wearable sensor design — low cost, low power, small footprint and inert skin interface. However, SCG can be challenging to process due to MEMS sensitivity to non-cardiopulmonary vibrations (vocalizations, motion artifacts, etc.) and varying waveform morphologies that challenge heuristics-based approaches.

### D. Time-Frequency Analysis in Sound

Short-time frequency analysis was established with the Fourier transform, the autocorrelation and the cepstrum defined and explored in the 1950s–1970s in strong collaboration with Bell Laboratories. [34] showed mathematically that STACFs should better represent signal periodicity over time than filter banks (later formalized as short-time Fourier transforms (STFTs) [35]). [36] and [37] showed how the short-time cepstral analysis could better isolate vocal pitch as the log Fourier transform of the vocal source is additive to that of the vocal tract in speech. [38] showed how subharmonic content could be consolidated in STFTs to help estimate pitch without cepstral analysis. With respect to fundamental rate detection and noise handling, voice frequency analysis is highly analogous to cardiopulmonary rate detection; however, the time scales and frequencies differ and methods were not unified. Little work has been done to bring short-time cepstral analysis to heart rate monitoring, and no cohesive framework has been assembled or benchmarked [39], [40].

### E. Contemporary HR and RR Estimation Approaches

Respiration rate and heart rate may be measured through many modalities including contact-based and non-contact-based mechanisms [16], [41], [42], [43]. Contemporary rate estimation approaches involve diverse preprocessing steps such as FIR, IIR, and adaptive filtering to de-noise signal and cancel artifacts. 1-D and 2-D signal transformations, decompositions such as the Shannon energy envelope, STFT, Empirical Mode Decomposition, and various wavelet transforms (e.g. the Continuous Wavelet Transform (CWT)) aim to separate signals of interest. Temporal and spectral methods typically involve peak finding heuristics to identify fiducial points. Exploration of 2-D signal transforms as neural network input remains an active area for research to address remaining unmet needs [5], [44]. Deep learning methods require large datasets to extrapolate to unseen populations and new signal sources beyond normative ranges. This data need can prohibit clinical deployment in unhealthy populations.

Until now, noninvasive fetal HR (fHR) from abdominal ECG (aECG) was a particularly challenging task due to the overlapping signal of interest from maternal and fetal HR. Blind source separation techniques involving ICA and PCA achieved limited success while template subtractions leave noisy, weak fECG that trouble temporal, spectral, wavelet, and machine-learning-based methods [1], [45], [46], [47], [48], [49]. Respiration monitoring in neonates is another clinically meaningful domain where current methods fall short. Standard-of-care uses wired bioimpedance (BioZ), which is cumbersome and difficult for staff to depend on for unhealthy neonates. Multiple frameworks have been presented to address the difficult task [50], [51].

## III. METHODOLOGY

CORAL estimates rates by generating a 2-D transformation over one or more time-discrete signals and finding an optimal traversal with Viterbi-style dynamic programming. We revisit **short-time autocorrelation functions** — a transformation over one time-discrete signal — and define the **correloform** — a reduction on one or more short-time autocorrelation functions. We define the correloform such that short-time autocorrelation functions may be referred to as a correloform.

### A. The Autocorrelation Function

Concisely, the autocorrelation function (ACF) is defined in the temporal domain as the correlation of a signal with itself for all valid offsets of the signal. For a signal, $x$ of length $N$, we can define the $\mathrm{ACF}(x)$ for each possible lag $i \in [0, N)$

$$\mathrm{ACF}(x)_i = \sum_{k=i}^{N-1} x_k x_{k-i} \tag{1}$$

If each $x_i$ is finite, the zero-lag value of the ACF is always maximal. The normalized autocorrelation (nACF) is obtained by dividing by the autocorrelation's zero-lag.

$$\mathrm{nACF}(x)_i = \frac{\mathrm{ACF}(x)_i}{\mathrm{ACF}(x)_0} \tag{2}$$

As $\mathrm{ACF}(x)_0$ is the sum of each element squared, the normalized scaling factor is proportional to the energy or average power, $P$, of the discrete-time signal.

$$P = \frac{\sum_{i=0}^{N-1} x_i^2}{N} \tag{3}$$

The nACF produces an output in $[-1, 1]$ such that 1 indicates perfect correlation, 0 indicates no correlation, and −1 indicates anti-correlation.

The autocorrelation has a desirable mathematical property that it may be computed via the frequency-domain [37], [52].

$$\mathrm{ACF} = \mathcal{F}^{-1}\left(|\mathcal{F}(x)|^2\right) \tag{4}$$

This formulation may be more computationally efficient to calculate as the time-domain formulation has $O(N^2)$ runtime and the frequency-domain formulation may be conditioned with $O(N \log N)$ runtime via choice of $N$.

Computation of Fourier transforms is well optimized, so the ACF may be implemented efficiently via the fast Fourier transform (FFT) and inverse fast Fourier transform (IFFT). After discretization and zero-padding to avoid circular convolutional artifacts

$$\mathrm{ACF} \approx \mathrm{IFFT}\left(|\mathrm{FFT}(x)|^2\right) \tag{5}$$

Fig. 1 illustrates nACF computed from a 3-second duration sample of ECG, PCG, and PPG. As lags are in units of samples, $i \in [0, N)$, the frequency associated with a lag is inversely proportional, so for a signal sampled at Fs, the frequency associated with $\mathrm{ACF}_i$ is $\mathrm{Hz}(i) = \frac{\mathrm{Fs}}{i}$ and $\mathrm{bpm}(\mathrm{lag}) = \mathrm{Hz}(\mathrm{lag}) * 60$.

The autocorrelation, autocorrelation function, and correlogram are used synonymously in various fields. It is worth mentioning an adjacent function with significant use in audio is the cepstrum [53]. In the context of the cepstrum, the lags are referred to as quefrencies with the inverse relationship to frequency.

$$\mathrm{C}_p(x) = \mathcal{F}^{-1}\left(\log\left(|\mathcal{F}(x)|^2\right)\right) \tag{6}$$

Because the logarithm of a product equals the sum of the logarithms of the factors, the autocorrelation is a better periodicity detector, while the cepstrum is a source separator [36]. Fig. 1 demonstrates the advantages of an autocorrelation over a cepstral approach.

### B. *Short-Time Autocorrelation Functions*

Short-Time Autocorrelation Functions (STACFs) differ from the spectrogram in that the $\mathrm{ACF}(x)$ is computed via $\mathrm{IFFT}(|\mathrm{FFT}(x)|^2)$ instead of $\mathcal{F}(x)$ via $\mathrm{FFT}(x)$ at each hop [34], [35].

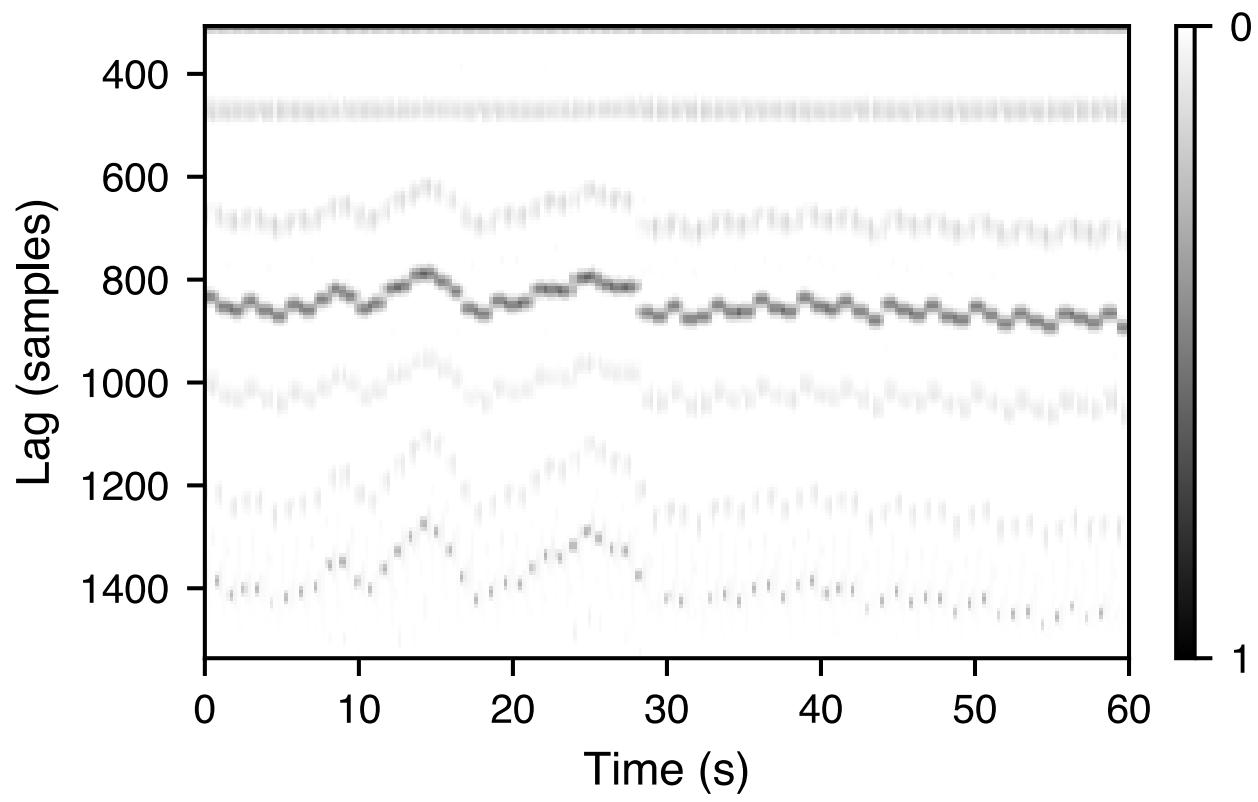

Fig. 2. A normalized Short-Time Autocorrelation Functions visualization of 1024 Hz ECG from a healthy adult male exhibiting 70–75 bpm HR, computed with 2048-sample autocorrelations and 128-sample hops.

Given a time-discrete signal $z$ of length $M$, STACFs produce a 2-D matrix of time steps (hops) by autocorrelation lags, parameterized by $N$, the length of the autocorrelation window, and $S$, the size of the hop. Therefore, the number of hops, $T$, is limited by the size of the autocorrelation window and the number of steps.

$$T = \begin{cases} 0 & \text{if } M - N < 0 \\ \left\lfloor \frac{M-N}{S} \right\rfloor + 1 & \text{otherwise} \end{cases} \tag{7}$$

Now, we may define the signal window under inspection to produce the columns $\mathrm{STACF}(z)$ over $t \in [0, T)$:

$$\forall j \in [0, N),\ y_t[j] = z[tS + j] \tag{8}$$

Each window is mean-centered before autocorrelation:

$$\forall j \in [0, N),\ \tilde{y}_{t[j]} = y_{t[j]} - \frac{1}{N}\sum_{m=0}^{N-1} y_{t[m]} \tag{9}$$

With the mean-centered windows, $\tilde{y}_t$, defined, we may write the short-time autocorrelation functions as

$$\begin{gathered} \forall t \in [0, T), i \in [0, N), \\ \mathrm{STACF}(z)_{ti} = \mathrm{ACF}(\tilde{y}_t)_i \end{gathered} \tag{10}$$

While the sample rate of $z$ is not a parameter of the STACF formulation, choice of $N$ such that window duration $\frac{N}{\mathrm{Fs}}$ spans one or more periods of interest is required to generate a response.

$\mathrm{nSTACF}(z)$ — the normalized transform scaled [−1, 1] with the same correlative interpretation as the $\mathrm{nACF}(x)$ at each hop — may be defined as:

$$\begin{gathered} \forall t \in [0, T), i \in [0, N) \\ \mathrm{nSTACF}(z)_{ti} = \mathrm{nACF}(\tilde{y}_t)_i \end{gathered} \tag{11}$$

When visualizing the nSTACF, we found it easier for interpretation if the domain is clamped [0, 1] such that white is ≤0 and black is 1. Fig. 2 illustrates how a nSTACF trivializes rate tracking in ECG signal with black indicating no and anti-correlation and white indicating strong periodicity. The repetition of the entire PQRST complex contributes to the dominant ridge with some less prominent ridges due to some positive correlation at lags aligning subcomponents of the complex like P waves and T waves and aligning in follow up beats.

### C. The Correloform

We define the **correloform** as a reduction over $K$ aligned STACFs of $T \times N$ dimensions with subharmonic summation $\alpha \in [0, 1]$. The raw $C$-channel signal $z$ has dimensions $M \times C$; with $W$ autocorrelation window sizes, $K = CW$ STACFs are reduced into $\mathcal{K}(z)$. Adaptive channel selection via mean-centered signal energy is implemented through the lens of this hop-wise energy normalization. As noted in Eq. (2) and Eq. (3), $\mathrm{ACF}(x)_0$ is proportional to the energy of the signal. Parameterized by $\alpha$, subharmonic summation consolidates support for the fundamental rate. With $K = 1$ and $\alpha = 0.0$, the correloform is the nonnegative nSTACF over the selected lag range.

$$\mathcal{K}'(z)_{ti} = \max\left(0, \frac{\max_k\left(\frac{\mathrm{ACF}(\tilde{y}_t^k)_i}{N_k}\right)}{\max_k\left(\frac{\mathrm{ACF}(\tilde{y}_t^k)_0}{N_k}\right)}\right) \tag{12}$$

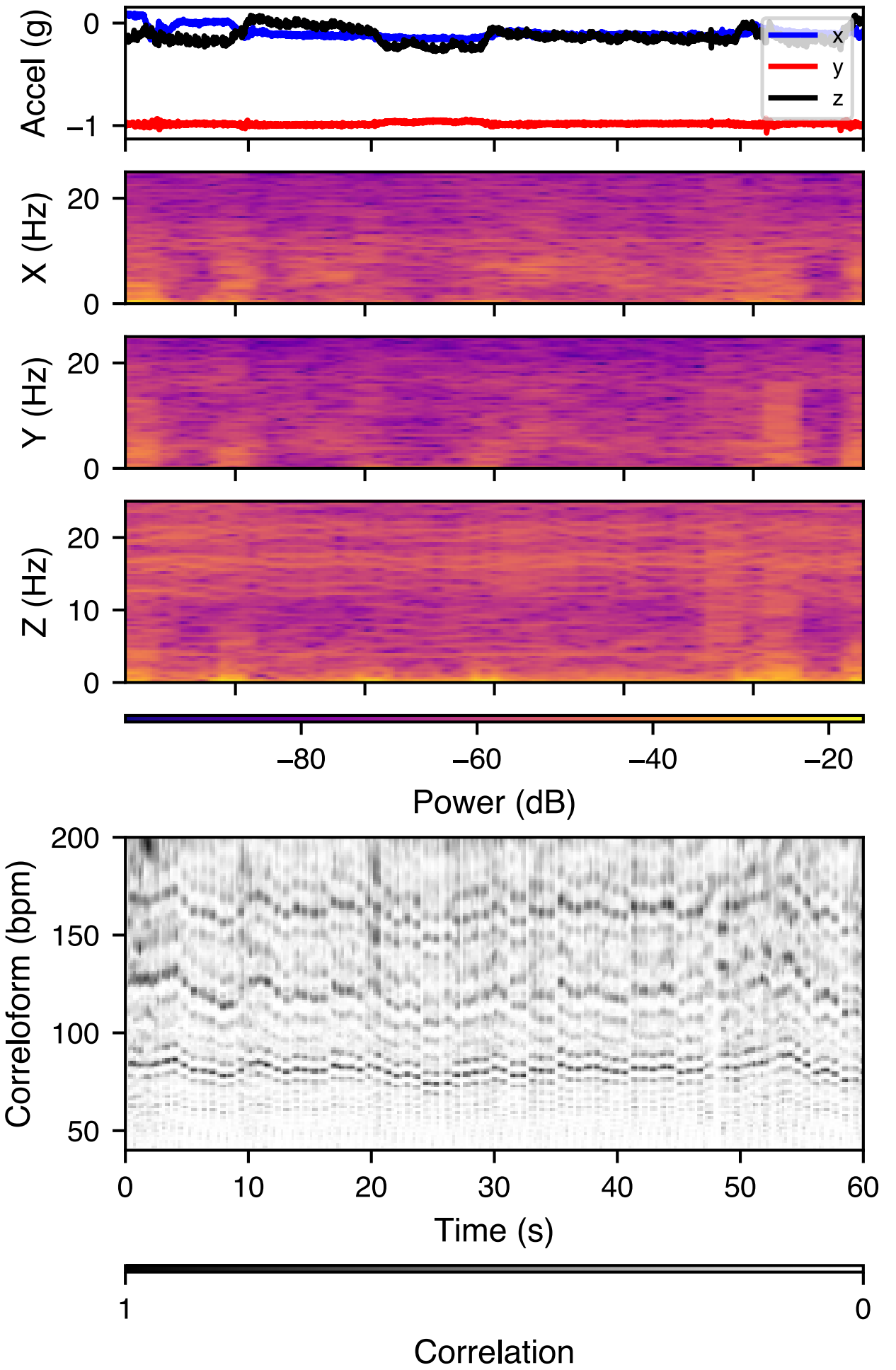

Fig. 3. Correloform and STFT analysis with the same 2-second FFT length on 1-minute of SCG HR — triaxial 1.6 kHz from suprasternal notch in supine.

For $\alpha \in (0, 1]$, support from the second- and third-multiple lags is added with weights $\alpha$ and $\alpha^2$; $\alpha = 0$ disables summation.

$$\mathcal{K}(z)_{ti} = \mathcal{K}'(z)_{ti} + \sum_{q=2}^{3} \alpha^{q-1} \max_{d \in \{-1,0,1\}} \mathcal{K}'(z)_{t(qi+d)} \tag{13}$$

Multiple autocorrelation window sizes may help to span slow and rapid changes in rate, morphology, and signal quality. If CORAL is used with multiple autocorrelation window sizes $N_k$, then not all STACFs will have the same $T_k \times N_k$ dimensions. Hops must be dropped from the edges and correlation values defaulted to $-\infty$ such that all $K$ STACFs have the same $T \times N$ dimensions prior to max aggregation.

Overall construction of the correloform from $z$ requires $O(KTN \log N)$ runtime.

### D. CORAL: Dynamic Programming and the Correloform

With the $\mathcal{K}(z)$ populated, extraction of the rate of interest is posed as a dynamic programming problem with similar structure to the Viterbi algorithm [54]. Successive rates are often similar, causing a ridge to track over time. We define several hyperparameters to allow tuning CORAL to the rate estimation application and quasi-periodic mechanism:

- $\beta$: The duration in seconds of the centered local-rate window.
- $\gamma$: The MAD multiplier for support-constrained replacement of local-rate outliers.
- $\delta$: The scaler to promote or demote detection of periodicity at each $\mathcal{K}(z)_{ti}$.
- $\varepsilon$: The scaler on penalty for square of percent bpm difference from median bpm for a hop.
- $\zeta$: The scaler on penalty for square of percent bpm change from one hop to the next.
- $\eta$: The even number of neighbor hops to include when reducing signal quality by neighbor correlation.

CORAL's rate extraction rests on a single core assumption, that the dominant periodicity is the complex of interest, with standard signal-conditioning applied per modality. At each hop $t$, CORAL first identifies the strongest correloform-supported lag, $w_t = \arg\max_j \mathcal{K}_{tj}$, with $r_t = \mathrm{bpm}(w_t)$, excluding the shortest 10% of lags to reduce high-frequency noise. Within the centered $\beta$-second window, the initial median over $r_q$ candidates is computed. When the ratio between successive candidate rates exceeds $1 + \frac{\Delta_{\mathrm{up}}}{100}$ or falls below $1 - \frac{\Delta_{\mathrm{down}}}{100}$, the candidate farther from the window median is replaced by the strongest correloform-supported rate at that hop within the corresponding asymmetric band about the median. For all reported benchmarks, $\Delta_{\mathrm{up}} = \Delta_{\mathrm{down}}$; Table S2 reports their common value as $\max\Delta\%$. The median and MAD are then recomputed, candidates beyond $\gamma \times \mathrm{MAD}$ undergo the same support-constrained replacement, and $f_t$ is the final window rate estimate. Dynamic programming retains the full configured search range.

Let $Q$ be the number of integer lag states in the configured rate-search interval. We define a $T \times Q$ reward/cost matrix $R$ that optimizes the sum of rewards for visiting high periodicity and penalties for changing rates and deviating from the local median rate. For $t = 0$, $R_{ti} = \mathcal{K}(z)_{ti}$. $R$ is efficiently memoized successively from $t = 0$ to $t = T - 1$ with correlation reward ($a$), fundamental rate deviation penalty ($b$), and rate change penalty ($c$). For a rate search, e.g. $\min = \frac{30 \text{ bpm}}{60 \text{ bpm}}$ and $\max = \frac{220 \text{ bpm}}{60 \text{ bpm}}$, we use $j \in \left[\frac{\text{Fs}}{\max}, \frac{\text{Fs}}{\min}\right]$. A second $T \times Q$ array, $L$, holds the predecessor lag index, $j$.

$$
\begin{aligned}
a &= \delta \mathcal{K}(z)_{ti} \\
b &= \varepsilon \left(100 * \frac{\text{bpm}(i) - f_t}{f_t}\right)^2 \\
c &= \zeta \left(100 * \frac{\text{bpm}(i) - \text{bpm}(j)}{\text{bpm}(i)}\right)^2 \\
R_{ti} &= \max_j \Big(R_{(t-1)j} + a - b - c\Big) \\
L_{ti} &= \arg\max_j \Big(R_{(t-1)j} + a - b - c\Big)
\end{aligned}
\tag{14}
$$

$L$ is traversed in reverse to follow the optimal path $P$ from $t = T - 1$ to $t = 0$. For the base case $t = T - 1$, the chosen lag is $P_t = \arg\max_j R_{tj}$. For each successive $t$,

$$
P_t = L_{t(P_{t+1})} \tag{15}
$$

The signal quality index (SQI) is indicated by the normalized periodicity and temporal similarity via $\mathcal{K}(z)_{tP_t}$ and Pearson's $r$ between adjacent hops, $\rho(\cdot, \cdot)$. By the same core assumption, estimates are likely noise if a short window of neighbors shows dissimilar periodicity or the path does not visit strong correlation values.

$$
\begin{aligned}
\text{SQI}_t = \mathcal{K}(z)_{tP_t} & \\
&+ \frac{\sum_{i=1}^{\frac{\eta}{2}} \Big(\mathcal{K}(z)_{(t-i)P_{t-i}} + \mathcal{K}(z)_{(t+i)P_{t+i}}\Big)}{\eta} \\
&\times \frac{\sum_{i=1}^{\frac{\eta}{2}} \rho\big(\mathcal{K}(z)_t, \mathcal{K}(z)_{t-i}\big) + \rho\big(\mathcal{K}(z)_t, \mathcal{K}(z)_{t+i}\big)}{\eta}
\end{aligned}
\tag{16}
$$

At record edges, unavailable neighbors are omitted and each sum is divided by the number available; without neighbors, $\text{SQI}_t$ is the center periodicity.

The centered local-rate window contains at most $B = 2\left\lceil \frac{\beta \, \text{Fs}}{2S} \right\rceil + 1$ hops. The dominant-lag scan costs $O(TQ)$. Median and MAD computation costs $O(TB)$; support-constrained replacement is at most $O(TBQ)$. The dynamic programming algorithm has worst-case runtime $O(TQ^2)$. The SQI reduction requires $O(T(\eta + 1)Q)$. Dynamic programming dominates when $Q \gg B$ and $Q \gg \eta + 1$.

a) *Processing Pipeline:*

CORAL solely assumes that the signal of interest represents the dominant periodicity and extracts the dominant rates over time that make up the input signal. Each biosignal is sampled such that there is a sub-frequency range of interest that contains the most interesting signal to analyze and some additional noise. The usual steps to prepare some signal for CORAL include IIR/FIR filtering, linear interpolation to a power-of-two frequency, and generation of $\mathcal{K}(z)$. Dataset-specific signal conditioning/masking and CORAL parameters are reported in Tables S1 and S2, respectively.

For fHR from aECG, mQRS complexes are detected from a thoracic maternal channel when available and otherwise from aECG. Each abdominal lead undergoes per-beat least-squares mQRS template subtraction using the 2 preceding and 2 following mQRS complexes. The residual is Gaussian highpass filtered to 10 Hz and converted to its Hilbert envelope. Channels are ranked by mean 1 s crest factor (CF). In FSYNTH, high-RMS spans lasting at least 30 s are automatically masked with a 5 s guard against uterine-contraction EMG contamination.

Coverage is reported as CORAL time divided by reference-comparable time; unavailable or invalid reference intervals are outside the denominator, whereas signal masks, missing CORAL output, and SQI rejection, where applied, reduce coverage. Agreement is the percentage of reported estimates with $|\text{CORAL} - \text{reference}| \leq \max(5 \ \text{bpm}, 0.10 \ |\text{reference}|)$ for HR/fHR or $\leq \max(2 \ \text{brpm}, 0.10 \ |\text{reference}|)$ for RR. Benchmark-specific reporting rules, rate processing, coverage, $r$, and agreement are reported in Table S3.

## IV. DATASETS AND RESULTS

We evaluate the performance of CORAL on a wide breadth of benchmarking and wearable datasets. Each dataset provides a unique aspect to the evaluation such as new age groups, diseases, modalities, device(s), and acquisition setting.

### A. *Publicly Available Benchmarking Datasets*

As CORAL is specific to neither morphology nor modality, we include benchmarking datasets to establish baseline capabilities against previously published work.

a) *M4WDB: Continuous Bedside ECG (N = 197):*

M4WDB contains continuous ICU ECG from MIMIC-IV-WDB [55], [56]. Figs. S18–S20 compare instantaneous CORAL and NeuroKit2 ECG HR, contextualize Pan–Tompkins disagreement, and report HRV.

b) *PTB-XL: Diagnostic 12-Lead ECG (N = 18,868):*

Well-supervised and containing a large proportion of healthy controls, PTB-XL's diagnostic 12-lead ECG provides further insight into cardiovascular diagnostic superclasses [57]. Figs. S6 and S7 compare CORAL HR and HRV with Pan–Tompkins; Fig. S8 compares detector RMSE and CORAL SQI across diagnostic superclasses.

Table I. CORAL HR (bpm) Performance Across Datasets

| Dataset | Modality (Channels) | $r$ | RMSE | Bias (±LoA) |
|---|---|---|---|---|
| LABOR | aECG (4) | 0.995 | 0.7 | 0.0 ± 1.5 |
| FSYNTH | aECG (32) | 0.984 | 3.5 | 0.1 ± 6.8 |
| CEBS | ECG (1) | 0.995 | 1.2 | 0.0 ± 2.3 |
| PTB-XL | ECG (12) | 0.968 | 4.4 | 0.1 ± 8.6 |
| M4WDB | ECG (1–7) | 0.868 | 14.1 | 1.2 ± 27.5 |
| ARC | ECG (1) | 0.970 | 3.4 | 0.5 ± 6.6 |
| ASXCG | ECG (1) | 0.990 | 1.5 | 0.0 ± 3.0 |
| SLEEP | ECG (1) | 0.952 | 5.1 | 0.6 ± 10.0 |
| SLEEP | PPG (1) | 0.971 | 3.9 | −0.3 ± 7.7 |
| NIRS | PPG (8) | 0.974 | 1.9 | 0.0 ± 3.7 |
| ASXCG | PCG (1) | 0.972 | 2.6 | 0.1 ± 5.2 |
| SLEEP | PCG (1) | 0.958 | 4.8 | 1.0 ± 9.2 |
| ASXCG | SCG (1) | 0.960 | 3.1 | 0.1 ± 6.1 |
| CEBS | SCG (1) | 0.990 | 1.6 | 0.0 ± 3.1 |
| AMB | SCG (3) | 0.970 | 4.9 | 0.8 ± 9.4 |
| MOUSE | ECG (1) | 0.934 | 32.6 | 1.8 ± 63.8 |

c) *CEBS: Combined Measurement of ECG, Breathing, and SCG (N = 20):*

Many SCG-derived HR and HRV works benchmark against CEBS [12], [58], [59], [60]. We compare annotated HR and HRV against CORAL on ECG and monaxial SCG (Figs. S2–S5) and peak-detection thoracic piezoresistive band (PRB) RR against CORAL PRB RR (Fig. S27), after removing duplicate recordings.

d) *LABOR: Abdominal and Direct Fetal ECG (N = 5):*

ADFECGDB contains noninvasive 4-channel abdominal ECG and invasive direct fetal ECG from 5 women in labor [45], [46]. CORAL estimates fHR from abdominal ECG after maternal-QRS subtraction. Direct fetal scalp ECG is used to derive reference RR intervals. We use the 3 highest-CF abdominal channels without contraction masking or SQI rejection. We compare CORAL fHR using LABOR-optimized parameters (Fig. S24) and the complete FSYNTH-optimized CORAL parameter set (Fig. S26).

e) *FSYNTH: Fetal ECG Synthetic Database (N = 10):*

FECGSYNDB simulates 32-lead aECG with 2 thoracic channels for maternal reference [47]. We evaluate 1,500 5-minute single-fetus records from 10 simulated subjects, excluding the twinning case. Fig. S25 reports instantaneous fHR with automatic high-RMS exclusion and across all time.

### B. Research Device Datasets

We feature several modalities of sensing from multiple medical research devices from the Rogers research group. Collection scenarios cover at-home, in-clinic, and in-vivo. Comparisons are made against an established detector or a secondary modality such as an internal, single-lead ECG or external gold-standard machinery when available.

a) *ASXCG: Aortic Stenosis Screening (N = 39):*

Table II. CORAL RR (brpm) Performance Across Datasets

| Dataset | Modality (Channels) | $r$ | RMSE | Bias (±LoA) |
|---|---|---|---|---|
| CEBS | PRB (1) | 0.992 | 0.5 | 0.3 ± 0.9 |
| SLEEP | RIP (2) | 0.966 | 1.2 | 0.1 ± 2.3 |
| SLEEP | SCG (3) | 0.946 | 1.4 | 0.1 ± 2.8 |
| SLEEP | Airflow (1) | 0.909 | 2.0 | −0.1 ± 3.9 |
| ARC | BioZ (1) | 0.858 | 8.1 | 1.9 ± 15.3 |

From the chest, 5-minute sessions from patients with mild to severe aortic stenosis produce many ectopic beats and elongated S1 to S2 heart sound intervals and lower SNR that trouble heuristics in automated interval, template, and frequency based algorithms for HR in vibrocardiography [61]. We evaluate continuous ECG-derived HR against CORAL HR on ECG, PCG, and SCG in Figs. S9–S11.

b) *ARC: Premature Neonatal Monitoring (N = 6):*

Six premature neonates with severe cardiopulmonary disease were monitored in the NICU using the wireless ARC wearable, which recorded single-lead ECG and BioZ (Kwak et al., provisionally accepted at *Device*). CORAL HR from ARC ECG was compared with Pan–Tompkins HR from the same ECG (Fig. S21), and CORAL RR from ARC BioZ was compared with breath-interval RR calculated from the simultaneously recorded wired Philips impedance waveform (Fig. S22).

c) *SLEEP: Overnight Sleep Studies (N = 10):*

Pediatric sleep study subjects completed overnight polysomnography (PSG) with wearable PCG for extended HR and RR comparisons [62]. Across typical sleep-study artifacts, CORAL HR from ECG, PCG, and PPG is compared with clinical PSG HR in Figs. S15–S17. CORAL RR from RIP and SCG is compared with peak-detection RIP RR, and CORAL airflow RR is compared with thermistor breath-interval RR, in Figs. S28–S30.

d) *NIRS: Multi-nodal NIRS PPG (N = 1):*

With an integrated ECG/PCG and 1-, 4-, or 8-channel NIR PPG at each hand and foot during change in body orientation and Valsalva maneuvers, NIRS produces rapid blood pressure and HR change (Hua et al., manuscript in preparation). Automatic channel selection is required as some photodetectors saturate and a blood pressure cuff periodically reduces circulation to one sensor. Continuous CORAL PPG HR is compared with NeuroKit2 HR in Fig. S12.

e) *AMB: Naturalistic Ambulatory SCG (N = 1):*

Using the hardware platform from [33], AMB is a normative ambulatory triaxial SCG dataset with a commercial chest-mounted ECG for reference. Figs. S13 and S14 compare continuous CORAL SCG HR with chest-strap ECG HR across the ~10 hour dataset, including ~1 hour of driving, ~1 mile of walking, ~8 hours of office work and meetings, and a 30-minute high-intensity stationary cycling workout.

f) *MOUSE: Wireless ECG Implant ($N = 1$):*

A wireless, battery-free murine implant recorded synchronous ECG over 24 hours in a naturalistic enclosure (Wang et al., manuscript in preparation). We compare CORAL HR against ECG envelope peak-detection HR; Fig. S31 demonstrates CORAL's domain-agnostic capabilities.

## V. DISCUSSION

In this work, we make heavy use of Bland–Altman analysis to reveal bias, limits of agreement, and agreement over the domain [63]. With diverse domains for comparison of CORAL rates, we also report Pearson's $r$ to indicate bidirectional linear performance and RMSE to indicate absolute agreement. Table S3 summarizes benchmark results, Fig. S1 defines the common density scale, and Figs. S2–S31 provide detailed comparisons. Per-subject Bland–Altman statistics (Table S4) confirm pooled agreement is not driven by between-subject spread. CORAL performs exceptionally well across fetal, neonatal, pediatric, adult, and geriatric cohorts for HR and RR estimation across clinical, wearable, and implantable datasets.

Nearly all benchmarks achieved $r > 0.95$. The two lowest correlations were both ICU datasets (M4WDB ICU ECG HR, $r = 0.868$, and ARC BioZ RR in premature neonates with severe cardiopulmonary disease, $r = 0.858$). In both ICU cases, CORAL met or exceeded previously published performance. Upon manual inspection, CORAL often correctly captured the dominant complex periodicity in the comparative biosignal while the reference detector may have been affected by abnormality or noise. As a fair comparison to reference signal and algorithms, where an SQI threshold was enforced, we increased the threshold until segments with artifacts were excluded or until SQI reached the median.

### A. Gold-Standard Algorithm Issues

Reference signals and algorithms did not always represent the periodicity present in the compared biosignal. Neonatal and pediatric reference data were particularly difficult to reduce because of motion artifacts, line noise, disconnections, and other contamination. Fig. S23 shows periodic breathing in ARC BioZ and acceleration that is not dominant in the simultaneous Philips impedance reference. Bigeminy causes an extra QRS complex in rapid succession; Fig. S19 demonstrates pulsatility recovered rather than heart rate in some cases.

### B. Robustness and Failure Modes

When the signal of interest is not the most powerful signal in the frequency band of interest, then by definition the correloform will be rewarded for tracking signals not of interest. Traditional domain-specific measures of SNR can aid signal preprocessing and masking. Among fHR signals retained by the automatic exclusion criteria, residual envelopes with a maximum crest factor of at least 3 produced high instantaneous rate agreement.

Of note, if $\zeta$ is too large, the traversal may begin to ignore real fluctuations like those due to RSA. RSA is caused by the natural inhibition of the onset of the heartbeat caused by vagal withdrawal during the expansion of the diaphragm in the inhalation portion of the respiratory cycle [64]. In a sequence of interbeat intervals, a shorter interbeat interval appears irregularly in a sequence of longer beats. By choosing an autocorrelation window that spans fewer heartbeat cycles, the autocorrelation can surface the rate corresponding to the locally abnormal beat interval due to RSA. By using short, medium, and longer autocorrelation windows, CORAL gracefully handles the localized rate due to RSA and the average rate between RSA. For this reason, the HR benchmarks used 0.5 s, 1 s, 2 s, and 3 s autocorrelation windows. Despite the lack of QRS detection, CORAL HR from both ECG and SCG correlates with interval-based SDNN and RMSSD in Figs. S3 and S5. With a smaller choice of $\zeta$, CSDNN and CRMSSD (CORAL-derived SDNN and RMSSD) are competitive with state-of-the-art on CEBS [58], [65].

Tracking complexes with multiple activations as in SCG is a particular strength of CORAL. In Fig. 3, we may observe the subject HR ($\approx$ 80 bpm) along the black dotted line. This circumstance illustrates both the automatic reduction of triaxial SCG and the inability of the STFT to accurately visualize the periodicity of the heartbeat complex in non-sinusoidal biosignals. A weakness of CORAL is a sustained asymmetric pattern in complexes cycle to cycle such as bimodal HR or alternating morphology. Subharmonic summation is tunable via $\alpha$ closer to 1 to better handle skip-a-beat or alternating morphology patterns that cause strong periodicity at half the fundamental rate. For noisier signals like SCG or fECG, $\alpha$ may be better closer to 0.

a) *Fetal HR from Abdominal ECG:*

Abdominal ECG is measured for the purposes of noninvasive fetal heart rate monitoring. Fig. 4 demonstrates maternal-QRS subtraction followed by instantaneous fetal-rate tracking in LABOR. LABOR had substantially higher residual fECG SNR than FSYNTH and required neither contraction exclusion nor SQI rejection. In FSYNTH, a synchronized thoracic maternal channel supported mQRS detection in the mixed-quality simulations. We recommend a synchronized maternal reference for aECG wearables intended to report fHR.

For instantaneous fHR results (Table S3; Figs. S24–S26), each eligible reference RR interval is scored at its midpoint. Allowing each bounding fQRS a ±50 ms tolerance gives the allowable rate band $\left[\frac{60}{\mathrm{RR}+0.1\ \mathrm{s}}, \frac{60}{\mathrm{RR}-0.1\ \mathrm{s}}\right]$. F1 is the fraction of supported eligible intervals within this band; F1 and coverage use interval counts. Cao et al. reported FSYNTH F1 $= 0.98890$ at 0 dB using 4 selected abdominal channels and 1 thoracic maternal channel [66]. Yang et al. reported F1 $=$ 0.9933 on ADFECGDB [67]. CORAL achieved F1 / coverage 1.000 / 99.1% on LABOR, 0.980 / 90.3% for FSYNTH with

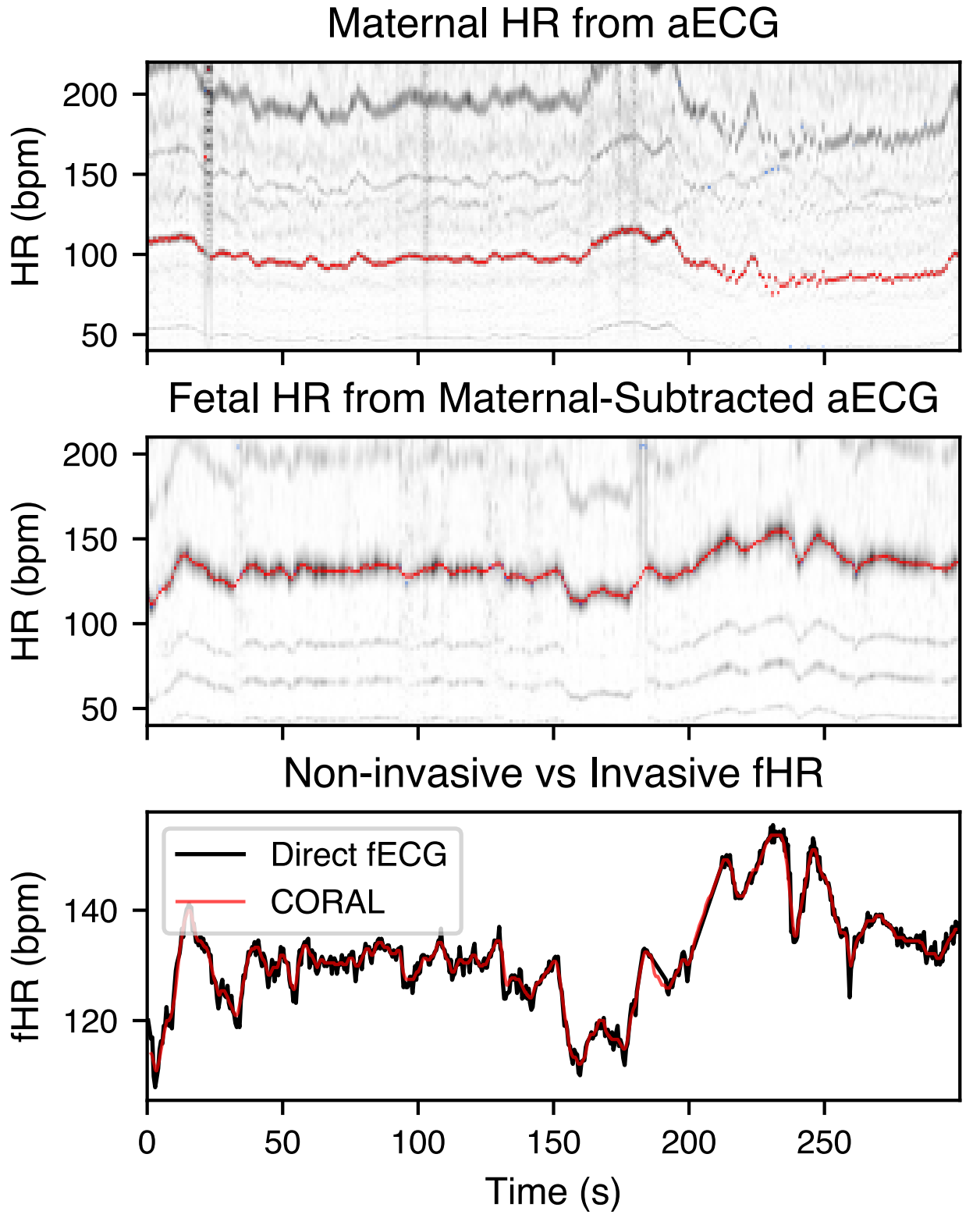

Fig. 4. CORAL HR for subject r10 from the LABOR dataset reveals maternal and fetal HR after maternal QRS subtraction. Comparison to direct fECG shows 99.0% coverage with tight agreement $r = 0.993$, $\mu = 0.08$ bpm (95% LoA $[-2.01, 2.17]$ bpm), and MAE $= 0.81$ bpm.

contractions, and 0.999 / 83.3% for FSYNTH after automatic EMG contamination exclusion.

b) *Heart Rate Variability:*

With CORAL's robust, granular estimate of heart rate, its variability correlates with interval-based measures of HRV. Over 60-second CEBS windows, ECG-derived CORAL HR SD correlated with annotated SDNN ($r = 0.759$), and CORAL HR RMSSD correlated with annotated RMSSD ($r = 0.725$; Fig. S3). The corresponding SCG-derived measures also correlated with annotated SDNN ($r = 0.733$) and RMSSD ($r = 0.643$; Fig. S5). These relationships persist at scale: over 60-second windows, CORAL ECG HR SD and RMSSD track the corresponding detector-derived measures in the M4WDB ICU cohort ($r = 0.768$ and 0.749; Fig. S20) and in 10-second PTB-XL strips ($r = 0.715$ and 0.663; Fig. S7). RMSSD agreement in the large continuous ICU dataset (0.749) is comparable to that in curated CEBS (0.725), suggesting the beat-to-beat variability estimate does not collapse in noisy real-world recordings.

c) *Future Work:*

Akin to neural networks, we believe the correloform has been neglected in modern research, perhaps because it was published in 1950 and the cepstrogram was later popularized and searchable. The autocorrelation recurs under many names across disciplines, from finance to medicine. Recent clinical work, for instance, derives postoperative-recovery indices from the autocorrelation of daily resting heart rate [10]. Convolutional neural networks remain a dominant pattern in medical machine learning yet struggle to represent periodicity and to count. Periodicity-aware 2D inputs remain an underexplored direction for convolutional and recurrent machine learning.

## VI. CONCLUSION

In this work, we define an easily interpretable signal transform to detect signal periodicity — the correloform — based on short-time autocorrelation functions and subharmonic summation. We define CORAL as a general purpose analytical framework encompassing the filtering, transformation, normalization, and traversal of signals for general, quasi-periodic rate and periodicity extraction. Through rigorous benchmarking, CORAL achieved robust HR and RR estimation without machine learning or assumptions about contour morphology or fiducial-point intervals across ECG, SCG, PCG, PPG, and BioZ from clinical and ambulatory settings and standard-of-care and research instruments.

CORAL demonstrates robust RR and HR estimation across all benchmarked populations, including automated noninvasive fetal HR, neonatal RR, and ambulatory SCG HR. CORAL achieves strong instantaneous fHR agreement on synthetic aECG against ground truth and clinical aECG against direct fECG. In premature neonates with severe cardiopulmonary disease, CORAL estimates RR from wearable BioZ with $r = 0.858$ against breath-interval RR from simultaneous wired Philips impedance. Although not designed for diagnosis, CORAL's signal quality index is systematically lower in normal than abnormal records across cardiovascular diagnostic superclasses (Holm-corrected p < 0.0001). Finally, CORAL's robust heart rate correlates strongly with annotated SDNN in healthy adults (CEBS ECG r=0.76 and SCG r=0.73), with moderate agreement for RMSSD.

## VII. ACKNOWLEDGMENT

The authors thank Professor John A. Rogers for his scientific guidance and institutional support and for fostering the interdisciplinary research environment in which this work was conducted.

The authors used OpenAI GPT-5.6 Sol and Anthropic Claude Opus 4.8 in an author-directed, human-in-the-loop workflow to aid correloform inspection and CORAL tuning in Section IV and the Supplemental Material.

## VIII. DECLARATIONS

**Data and code availability.** Public benchmark datasets are available through PhysioNet; other datasets are available upon author request. Upon publication, benchmark scripts will be available at github.com/trueb2/CORAL-2026, optimized Rust and portable Python implementations at github.com/trueb2/CORAL, and the Rust CLI at correloform.

**Ethical approval.** Only AMB involved new human data collection. The Northwestern University Institutional Review Board approved all AMB procedures, and the participant provided written informed consent (protocol STU00221958-CR0001). Remaining analyses used previously collected public or cited-study data, with ethical oversight reported by each source study.

**Funding.** CORAL development and analysis were supported by the Querrey Simpson Institute for Bioelectronics at Northwestern University; no external funding specifically supported this work.

# CORAL: A Modality-Invariant Framework for Robust Vital Sign Rate Estimation Using Correloform Analysis

*Supplementary Material*

Jacob Trueb, Amey Kasbe, Aniruddh Srinivasan

| Dataset | Signal | Rate | Conditioning | Masking and validity |
|---|---|---|---|---|
| CEBS | ECG | HR | 0.5–50 Hz bandpass | reference 30–220 bpm; exclude >30% adjacent RR change |
| CEBS | SCG | HR | 5–50 Hz bandpass | reference 30–220 bpm; exclude >30% adjacent RR change |
| PTB-XL | ECG ×12 | HR | 10–50 Hz bandpass | reference 20–300 bpm; require ≥4 reference leads |
| ASXCG | ECG | HR | 10–50 Hz bandpass | reference 25–220 bpm; exclude >30% adjacent RR change, ECG amplitude artifacts, and >1 s gaps |
| ASXCG | PCG | HR | 25–200 Hz bandpass | reference 25–220 bpm; exclude >30% adjacent RR change, ECG amplitude artifacts, and >1 s gaps |
| ASXCG | SCG | HR | 5–50 Hz bandpass | reference 25–220 bpm; exclude >30% adjacent RR change, ECG amplitude artifacts, and >1 s gaps |
| NIRS | PPG ×8 | HR | 0.5–10 Hz bandpass; first derivative | exclude flat channels; reference 30–220 bpm |
| AMB | SCG ×3 | HR | 18–40 Hz bandpass; 5 Hz energy envelope per axis | — |
| SLEEP | ECG | HR | 10–50 Hz bandpass | exclude 1 s QRS RMS <0.25× median; PSG HR 35–220 bpm; reference sample within 2 s |
| SLEEP | PCG | HR | 20–150 Hz bandpass | PSG HR 35–220 bpm; reference sample within 2 s |
| SLEEP | PPG | HR | 0.5–8 Hz bandpass; first derivative | exclude sensor dropout/saturation (1 s guard); PSG HR 35–220 bpm; reference sample within 2 s |
| M4WDB | ECG ×1–7 | HR | 8–25 Hz bandpass | exclude gaps, flatlines ≥0.5 s, and all-lead steps; require ≥1 valid lead; reference 25–220 bpm |
| ARC | ECG | HR | 10–50 Hz bandpass | reference 35–235 bpm before masking; exclude lead/contact alerts, low sampling, rails, gaps >1 s, and runs <10 s |
| ARC | BioZ | RR | 0.45–2.25 Hz bandpass; 0.2–2.25 Hz breath-interval consistency check | exclude off-body, low sampling, rails, flatlines ≥0.5 s, spikes (1 s guard), motion ≥100 mg, and runs <10 s |
| ARC | Philips BioZ (reference) | RR | 0.2–2.25 Hz bandpass | exclude low sampling, saturation, flatlines ≥2 s, and runs <10 s; reference 10–120 brpm |
| LABOR | aECG | fHR | 5–100 Hz bandpass; 60 Hz notch; ±2 mQRS subtraction; 10 Hz Gaussian highpass; Hilbert envelope | select top 3 channels; reference 40–210 bpm; exclude >30% adjacent change |
| FSYNTH | aECG | fHR | 3–124 Hz bandpass; 50/60 Hz notch; ±2 mQRS subtraction; 10 Hz Gaussian highpass; Hilbert envelope | crest factor within 1 of maximum; require maximum ≥3; reference 40–210 bpm; (optional) exclude ≥30 s high-RMS contractions (5 s guard) |
| CEBS | PRB | RR | 0.083–0.583 Hz bandpass | exclude breath-interval outliers; reference 5–35 brpm |
| SLEEP | RIP | RR | sum of separately median/MAD-normalized chest and abdomen; 0.08–0.70 Hz bandpass | exclude secondary peaks, breath-interval outliers, and ≥10 s low-amplitude RIP (<0.20× median); reference 5–72 brpm |
| SLEEP | SCG ×3 | RR | 0.10–0.70 Hz bandpass per axis; 20 s sliding PCA | exclude ≥10 s low-amplitude RIP (<0.20× median); reference 5–72 brpm |
| SLEEP | Airflow | RR | 0.03–0.70 Hz bandpass | exclude reference breaths failing amplitude, shape, continuity, or outlier checks and 30 s respiratory-to-wander SNR <0 dB; reference 5–72 brpm |
| MOUSE | ECG | HR | 20–120 Hz bandpass; Hilbert envelope | reference 200–1100 bpm; exclude gaps, runs <10 s, and >20% adjacent RR change |

Table S1: Systematic signal conditioning and masking.

| Dataset | CORAL band (Hz) | $\alpha$ | $\beta$ | $\gamma$ | $\delta$ | $\varepsilon$ | $\zeta$ | $\eta$ | Search ($min^{-1}$) | fs / hop | Windows | max$\Delta$% |
|---|---|---|---|---|---|---|---|---|---|---|---|---|
| CEBS ECG | 0.5–50 | 0.9 | 60 | 3 | 10 | 1e-3 | 0.01 | 4 | 25–220 | 512 / 128 | 256,512,1024,1536 | 40 |
| CEBS SCG | 5–50 | 0.0 | 60 | 3 | 10 | 1e-3 | 0.01 | 4 | 25–220 | 512 / 128 | 256,512,1024,1536 | 40 |
| PTB-XL | 10–50 | 0.5 | 10 | 3 | 100 | 1e-3 | 1e-3 | 4 | 25–220 | 512 / 128 | 256,512,1024,1536 | 40 |
| ASXCG ECG | 10–50 | 0.5 | 10 | 3 | 100 | 1e-3 | 1e-3 | 4 | 25–220 | 512 / 128 | 256,512,1024,1536 | 40 |
| ASXCG PCG | 25–200 | 0.0 | 60 | 3 | 10 | 1e-3 | 0.01 | 4 | 25–220 | 512 / 128 | 256,512,1024,1536 | 40 |
| ASXCG SCG | 5–50 | 0.0 | 60 | 3 | 10 | 1e-3 | 0.01 | 4 | 25–220 | 512 / 128 | 256,512,1024,1536 | 40 |
| NIRS PPG | 0.5–10 | 0.5 | 12 | 3 | 3 | 0.2 | 3.0 | 8 | 25–220 | 512 / 128 | 256,512,1024,1536 | 50 |
| AMB SCG | 0.5–4 | 0.0 | 60 | 3 | 10 | 0.0 | 5.0 | 4 | 25–220 | 512 / 128 | 512,1024,2048 | 60 |
| SLEEP ECG | 10–50 | 0.5 | 10 | 3 | 100 | 1e-3 | 1e-3 | 4 | 25–220 | 512 / 128 | 256,512,1024,1536 | 40 |
| SLEEP PCG | — | 0.0 | 60 | 3 | 10 | 1e-3 | 0.01 | 4 | 25–220 | 512 / 128 | 256,512,768 | 40 |
| SLEEP PPG | 0.8–8 | 0.5 | 10 | 3 | 100 | 1e-3 | 1e-3 | 4 | 25–220 | 512 / 128 | 256,512,1024,1536 | 40 |
| M4WDB | 8–25 | 0.25 | 10 | 3 | 100 | 1e-3 | 1e-3 | 4 | 25–220 | 512 / 128 | 256,512,1024,1536 | 40 |
| ARC ECG | 10–50 | 0.9 | 10 | 3 | 100 | 1e-3 | 0.1 | 4 | 25–235 | 512 / 128 | 256,512,1024,1536 | 40 |
| ARC BioZ | 0.45–2.25 | 1.0 | 5 | 2.5 | 10 | 1e-4 | 0.1 | 4 | 10–120 | 256 / 64 | 256,512,1024,2048 | 5 |
| LABOR | — | 0.6 | 10 | 3 | 10 | 1e-4 | 0.3 | 4 | 40–210 | 1024 / 256 | 512,1024,2048,3072 | 5 |
| FSYNTH | — | 0.6 | 10 | 3 | 10 | 1e-4 | 0.3 | 4 | 40–210 | 1024 / 256 | 512,1024,2048,4096,8192 | 5 |
| LABOR (FSYNTH parameters) | — | 0.6 | 10 | 3 | 10 | 1e-4 | 0.3 | 4 | 40–210 | 1024 / 256 | 512,1024,2048,4096,8192 | 5 |
| CEBS PRB | 0.083–0.583 | 0.0 | 30 | 3 | 5 | 1e-3 | 0.01 | 20 | 5–35 | 128 / 32 | 256,512,1024 | 15 |
| SLEEP RIP | 0.08–0.70 | 0.3 | 30 | 3 | 5 | 0.25 | 100 | 20 | 8–72 | 64 / 16 | 256,512,1024 | 15 |
| SLEEP SCG | 0.10–0.70 | 0.3 | 30 | 3 | 5 | 0.25 | 100 | 20 | 8–72 | 64 / 16 | 256,512,1024 | 15 |
| SLEEP Airflow | 0.03–0.70 | 0.3 | 30 | 3 | 5 | 0.25 | 100 | 20 | 8–72 | 64 / 16 | 256,512,1024 | 15 |
| MOUSE | — | 0.0 | 20 | 3 | 100 | 1e-3 | 0.1 | 4 | 200–1100 | 4096 / 512 | 512,1024,2048 | 40 |

Table S2: CORAL parameterization.

| Dataset | Reference | Reporting rule | Rate processing | Subjects | Coverage (N / M) | $r$ | Agreement (%) |
|---|---|---|---|---|---|---|---|
| CEBS ECG | annotated ECG HR | — | — | 20 | 18.28 / 18.39 h (99.4%) | 0.995 | 99.8 |
| CEBS SCG | annotated ECG HR* | — | — | 20 | 18.28 / 18.39 h (99.4%) | 0.990 | 99.1 |
| PTB-XL | Pan–Tompkins ECG HR (median ≥4 leads) | — | 10 s mean | 18,868 | 60.55 / 60.55 h (100.0%) | 0.968 | 96.2 |
| ASXCG ECG | Pan–Tompkins ECG HR | SQI ≥0.55 | — | 39 | 3.76 / 4.10 h (91.7%) | 0.990 | 99.1 |
| ASXCG PCG | Pan–Tompkins ECG HR* | SQI ≥0.25 | — | 39 | 3.35 / 4.10 h (81.6%) | 0.972 | 98.2 |
| ASXCG SCG | Pan–Tompkins ECG HR* | SQI ≥0.25 | — | 39 | 3.63 / 4.10 h (88.4%) | 0.960 | 97.6 |
| NIRS PPG | NeuroKit2 ECG HR* | SQI ≥10th percentile (0.68) | — | 1 | 0.41 / 0.46 h (89.9%) | 0.974 | 100.0 |
| AMB SCG | chest-strap ECG HR* | — | — | 1 | 10.18 / 10.18 h (100.0%) | 0.970 | 93.8 |
| SLEEP ECG | clinical PSG ECG HR | 10 s median SQI ≥0.6 | exponential mean (10 s half-life) | 10 | 87.73 / 90.54 h (96.9%) | 0.952 | 95.9 |
| SLEEP PCG | clinical PSG ECG HR* | 10 s median SQI ≥0.3 | exponential mean (10 s half-life) | 8 | 54.51 / 73.01 h (74.7%) | 0.958 | 95.7 |
| SLEEP PPG | clinical PSG ECG HR* | 10 s median SQI ≥1.0 | exponential mean (10 s half-life) | 4 | 32.24 / 36.01 h (89.5%) | 0.971 | 98.1 |
| M4WDB | NeuroKit2 ECG HR (lead median) | — | — | 197 | 8794.35 / 8805.49 h (99.9%) | 0.868 | 91.4 |
| ARC ECG | Pan–Tompkins ECG HR | — | 9 s mean | 6 | 92.54 / 101.91 h (90.8%) | 0.970 | 99.3 |
| ARC BioZ | Philips BioZ breath-interval RR* | 10-breath median breath-interval deviation ≤25% | 14-breath-interval 10% winsorized mean; 10-breath mean RR | 6 | 21.09 / 33.72 h (62.6%) | 0.858 | 78.2 |
| LABOR | fetal scalp ECG fHR* | — | — | 5 | 3,132 / 3,162 (99.1%) | 0.995 | 100.0 |
| FSYNTH | ground-truth fHR | — | — | 10 | 782,089 / 938,568 (83.3%) | 0.984 | 99.4 |
| FSYNTH (with contractions) | ground-truth fHR | — | — | 10 | 847,071 / 938,568 (90.3%) | 0.895 | 96.1 |
| LABOR (FSYNTH parameters) | fetal scalp ECG fHR* | — | — | 5 | 3,082 / 3,162 (97.5%) | 0.988 | 100.0 |
| CEBS PRB | PRB peak-detection RR | SQI ≥0.5 | 8 s median | 20 | 1.20 / 1.66 h (72.4%) | 0.992 | 99.7 |
| SLEEP RIP | RIP peak-detection RR | SQI ≥1.0 | 8 s median | 10 | 57.22 / 102.12 h (56.0%) | 0.966 | 97.4 |
| SLEEP SCG | RIP peak-detection RR* | SQI ≥1.0 | 8 s median | 8 | 35.14 / 74.28 h (47.3%) | 0.946 | 96.7 |
| SLEEP Airflow | thermistor breath-interval RR | SQI ≥1.0 | 8 s median | 10 | 38.88 / 63.16 h (61.6%) | 0.909 | 96.2 |
| MOUSE | ECG peak-detection HR (Kubios) | — | — | 1 | 22.21 / 22.21 h (100.0%) | 0.934 | 98.1 |

* Reference obtained from a separate measured signal.

Table S3: Systematic post-processing and benchmark results.

| Dataset | Subjects (LoA) | *N* | Pooled bias [LoA] | Subject bias median [IQR] | Subject bias range | Median subject LoA half-width |
|---|---|---|---|---|---|---|
| CEBS ECG | 20 (20) | 76,796 | +0.03 [−2.31, +2.37] | 0.00 [−0.02, +0.04] | [−0.09, +0.31] | 1.92 |
| CEBS SCG | 20 (20) | 76,796 | +0.04 [−3.08, +3.16] | −0.01 [−0.06, +0.05] | [−0.14, +0.46] | 2.41 |
| PTB-XL | 18,868 (2,111) | 21,798 | +0.09 [−8.55, +8.74] | +0.03 [−0.23, +0.35] | [−83.57, +77.05] | 0.73 |
| ASXCG ECG | 39 (39) | 15,072 | −0.02 [−3.01, +2.97] | 0.00 [−0.03, +0.01] | [−0.55, +0.29] | 0.67 |
| ASXCG PCG | 39 (39) | 13,476 | +0.08 [−5.10, +5.26] | −0.01 [−0.07, +0.02] | [−0.76, +6.03] | 1.62 |
| ASXCG SCG | 39 (39) | 14,570 | +0.11 [−5.98, +6.20] | 0.00 [−0.04, +0.03] | [−0.91, +8.96] | 1.36 |
| NIRS PPG | 1 (1) | 5,915 | −0.04 [−3.75, +3.68] | −0.04 [−0.04, −0.04] | [−0.04, −0.04] | 3.71 |
| AMB SCG | 1 (1) | 36,585 | +0.77 [−8.66, +10.20] | +0.77 [+0.77, +0.77] | [+0.77, +0.77] | 9.43 |
| SLEEP ECG | 10 (10) | 1,263,335 | +0.56 [−9.40, +10.52] | +0.38 [+0.10, +1.01] | [−0.51, +2.20] | 8.21 |
| SLEEP PCG | 8 (8) | 784,921 | +0.97 [−8.23, +10.18] | +1.08 [+0.83, +1.29] | [+0.27, +1.48] | 8.83 |
| SLEEP PPG | 4 (4) | 464,317 | −0.34 [−7.99, +7.32] | −0.28 [−0.69, +0.02] | [−1.35, +0.37] | 7.56 |
| M4WDB | 188 (188) | 126,638,679 | +1.22 [−26.24, +28.69] | +0.36 [+0.04, +1.85] | [−28.39, +76.95] | 17.50 |
| ARC ECG | 6 (6) | 1,332,562 | +0.49 [−6.08, +7.07] | +0.25 [+0.17, +0.39] | [−0.03, +1.09] | 6.61 |
| ARC BioZ | 6 (6) | 75,938 | +1.92 [−13.41, +17.25] | +1.94 [+0.23, +2.58] | [−2.27, +13.30] | 14.77 |
| LABOR | 5 (5) | 3,132 | +0.05 [−1.40, +1.50] | +0.04 [+0.03, +0.05] | [+0.03, +0.08] | 1.06 |
| FSYNTH | 10 (10) | 782,089 | +0.07 [−6.72, +6.86] | +0.07 [+0.01, +0.12] | [−0.08, +0.19] | 6.45 |
| FSYNTH (with contractions) | 10 (10) | 847,071 | +1.27 [−16.82, +19.37] | +1.48 [+0.77, +1.71] | [−0.53, +2.73] | 17.29 |
| LABOR (FSYNTH parameters) | 5 (5) | 3,082 | +0.06 [−2.11, +2.23] | +0.06 [+0.05, +0.13] | [−0.05, +0.13] | 1.70 |
| CEBS PRB | 19 (19) | 17,290 | +0.28 [−0.62, +1.18] | +0.27 [+0.25, +0.32] | [−0.04, +0.47] | 0.88 |
| SLEEP RIP | 10 (10) | 823,957 | +0.07 [−2.27, +2.41] | +0.07 [+0.04, +0.10] | [+0.01, +0.18] | 2.13 |
| SLEEP SCG | 8 (8) | 506,079 | +0.10 [−2.73, +2.93] | +0.07 [+0.04, +0.14] | [−0.09, +0.43] | 2.62 |
| SLEEP Airflow | 10 (10) | 559,853 | −0.06 [−4.00, +3.88] | +0.04 [−0.10, +0.09] | [−0.85, +1.09] | 2.83 |
| MOUSE | 1 (1) | 639,718 | +1.82 [−62.02, +65.65] | +1.82 [+1.82, +1.82] | [+1.82, +1.82] | 63.83 |

*Difference = CORAL − reference. HR/fHR values are bpm and RR values are brpm. Pooled estimates weight retained comparisons; subject summaries give each subject one value. Subject counts require ≥1 retained comparison; parentheses give subjects with estimable LoA (≥2 comparisons). IQR is the 25th–75th percentile. PTB-XL subjects usually contribute one 10 s strip; M4WDB is continuous, so their extreme ranges reflect single-strip and subject-level half/double-rate disagreements, respectively.*

Table S4: Per-subject Bland–Altman analysis.

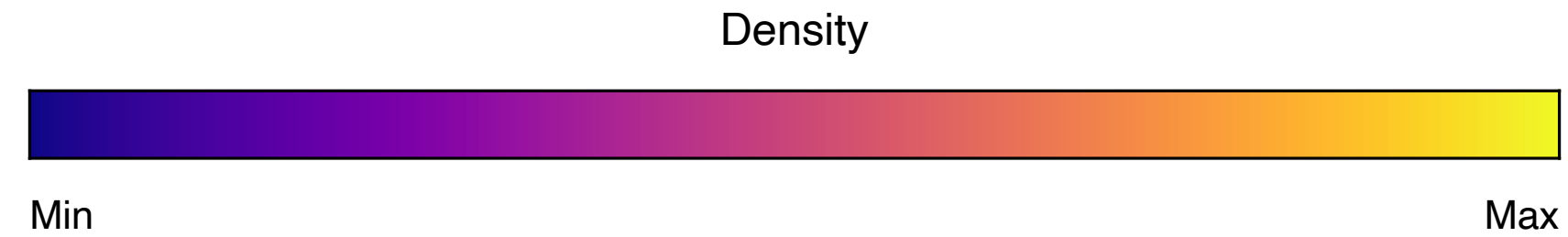

Figure S1: All supplemental density plots use this scale, independently normalized from minimum to maximum density.

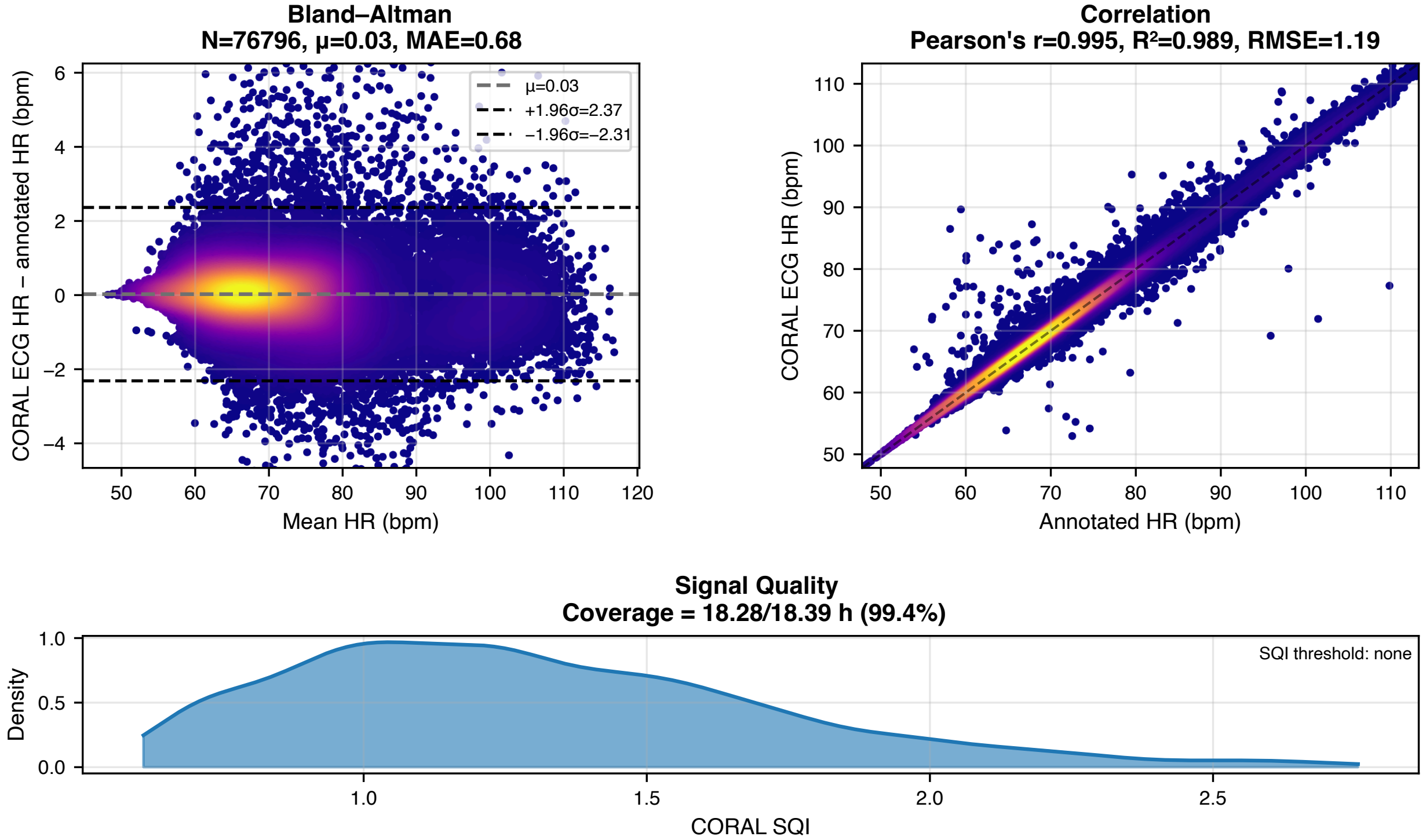

Figure S2: CORAL ECG HR performance against annotated ECG HR from CEBS.

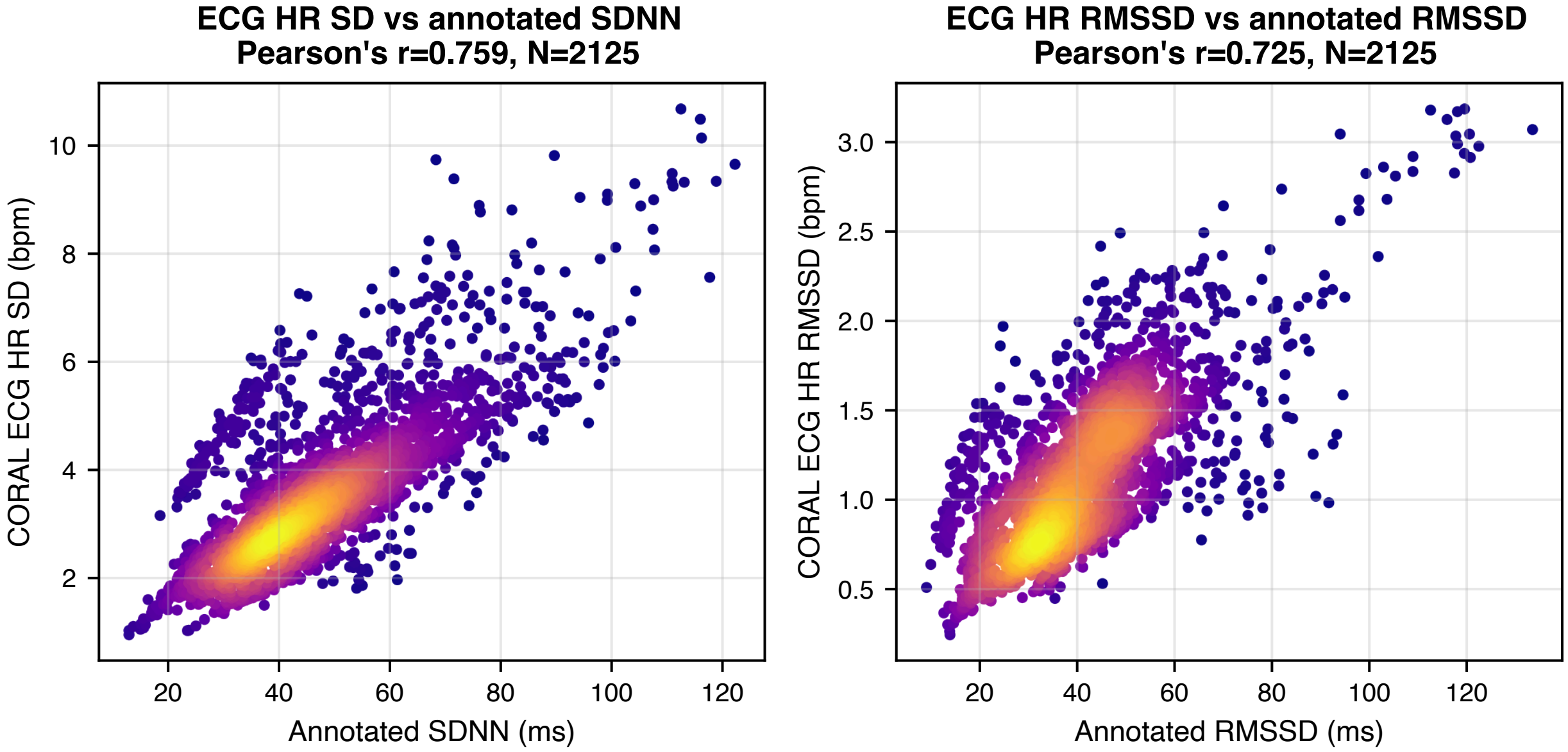

Figure S3: CORAL ECG HRV against annotated ECG HRV from CEBS (SD versus SDNN; RMSSD versus RMSSD).

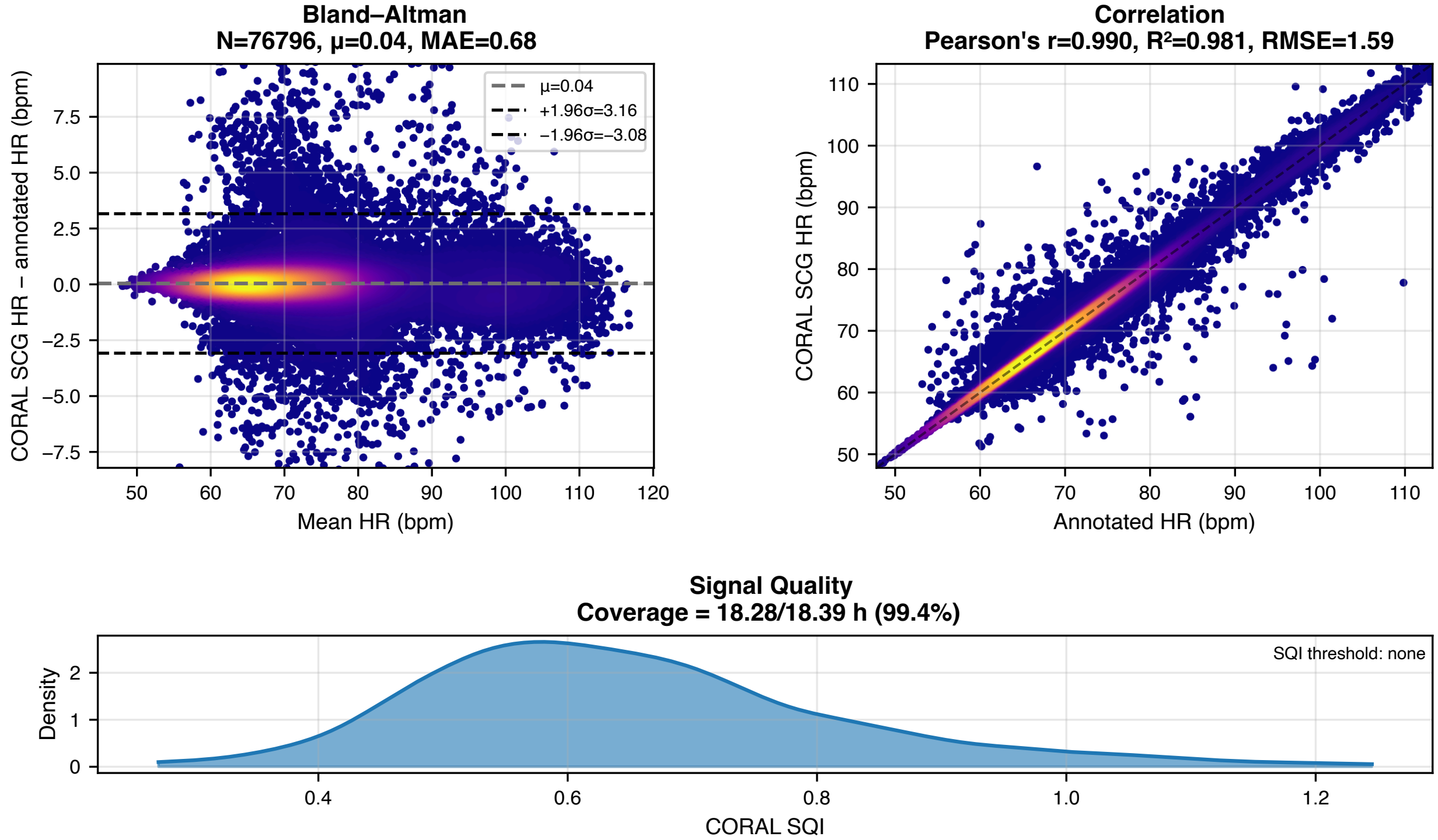

Figure S4: CORAL SCG HR performance against annotated ECG HR (separate reference signal) from CEBS.

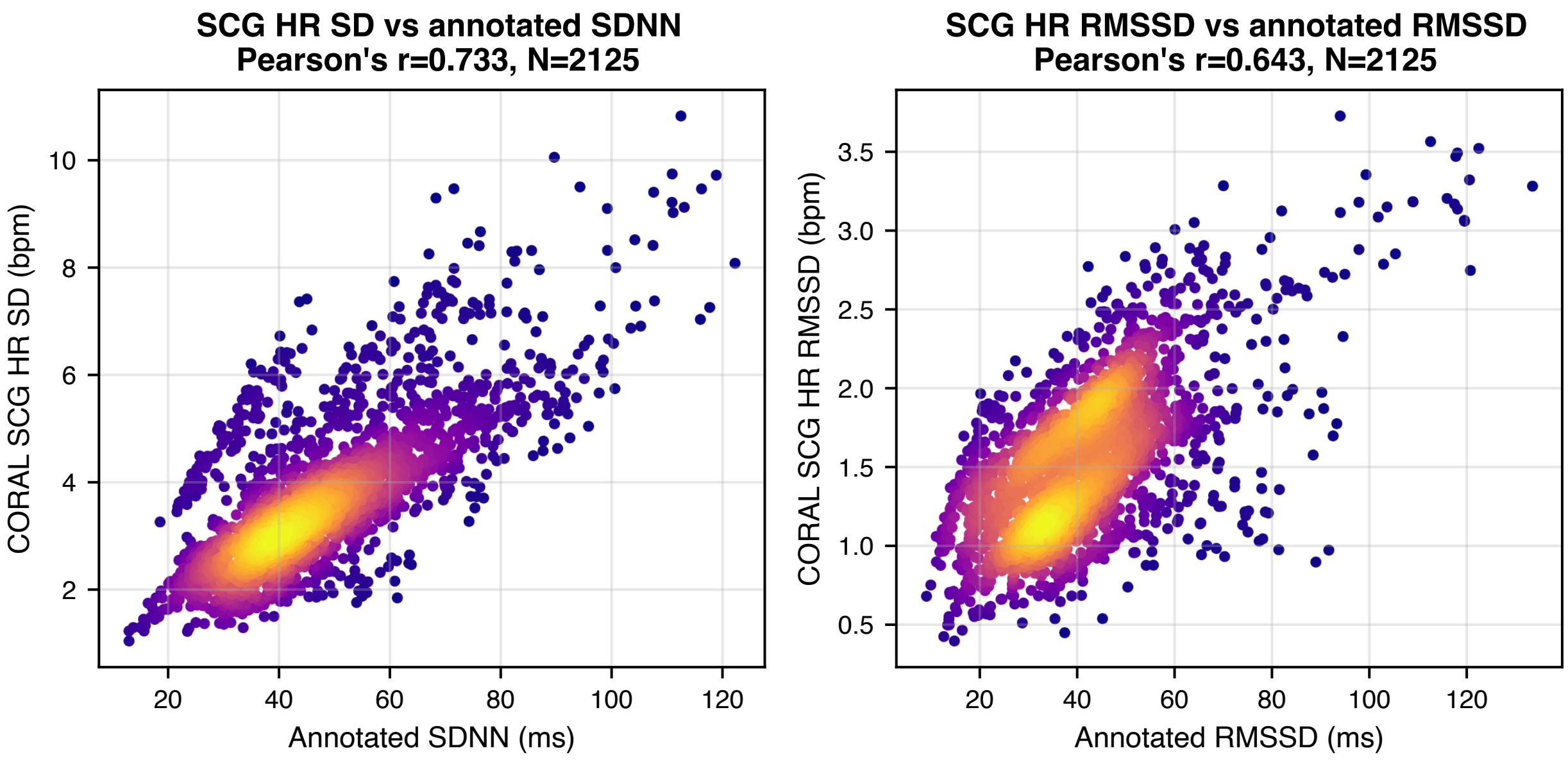

Figure S5: CORAL SCG HRV against annotated ECG HRV (separate reference signal) from CEBS (SD versus SDNN; RMSSD versus RMSSD).

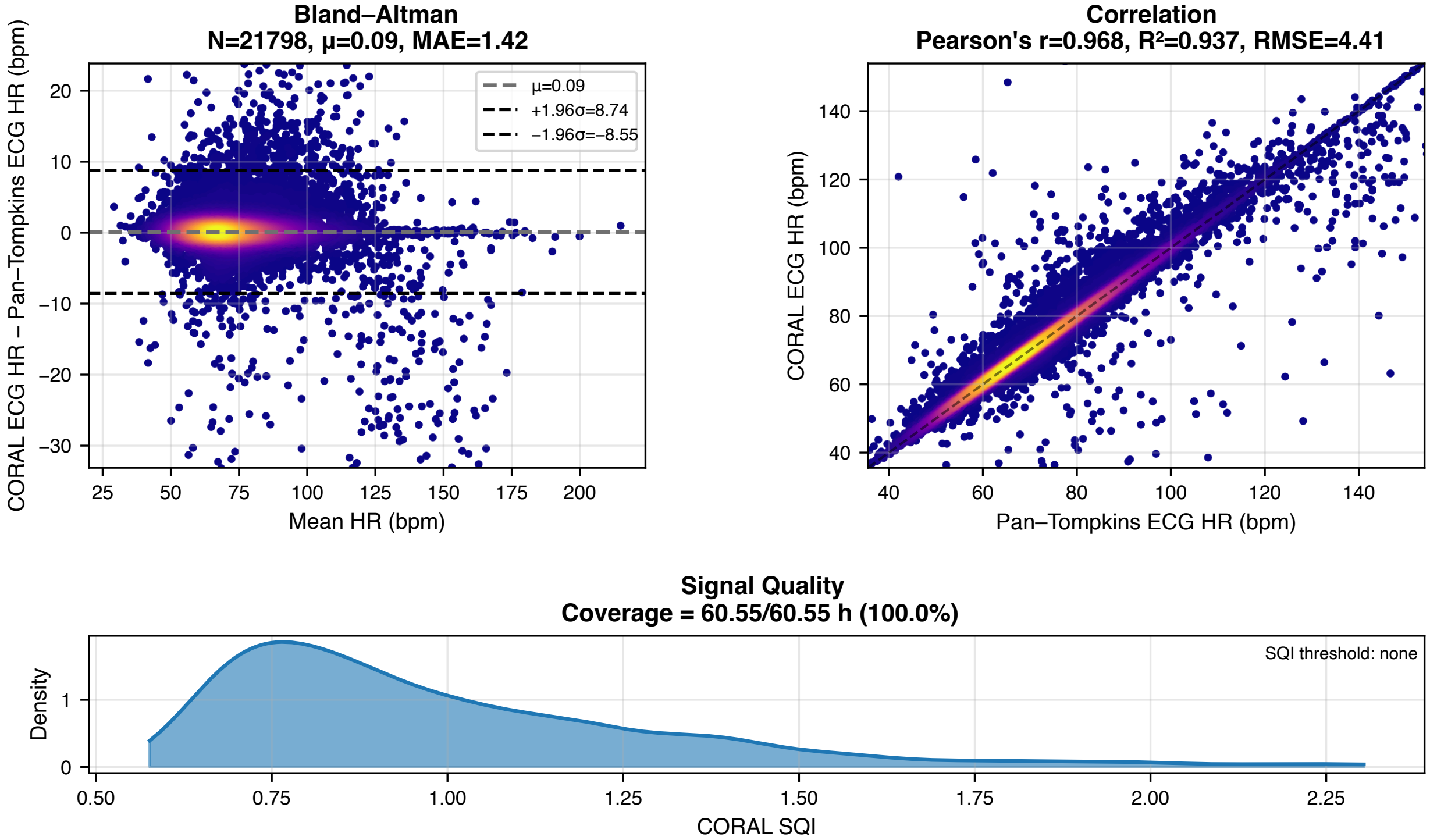

Figure S6: CORAL ECG HR performance against Pan–Tompkins ECG HR (reference detector) from PTB-XL.

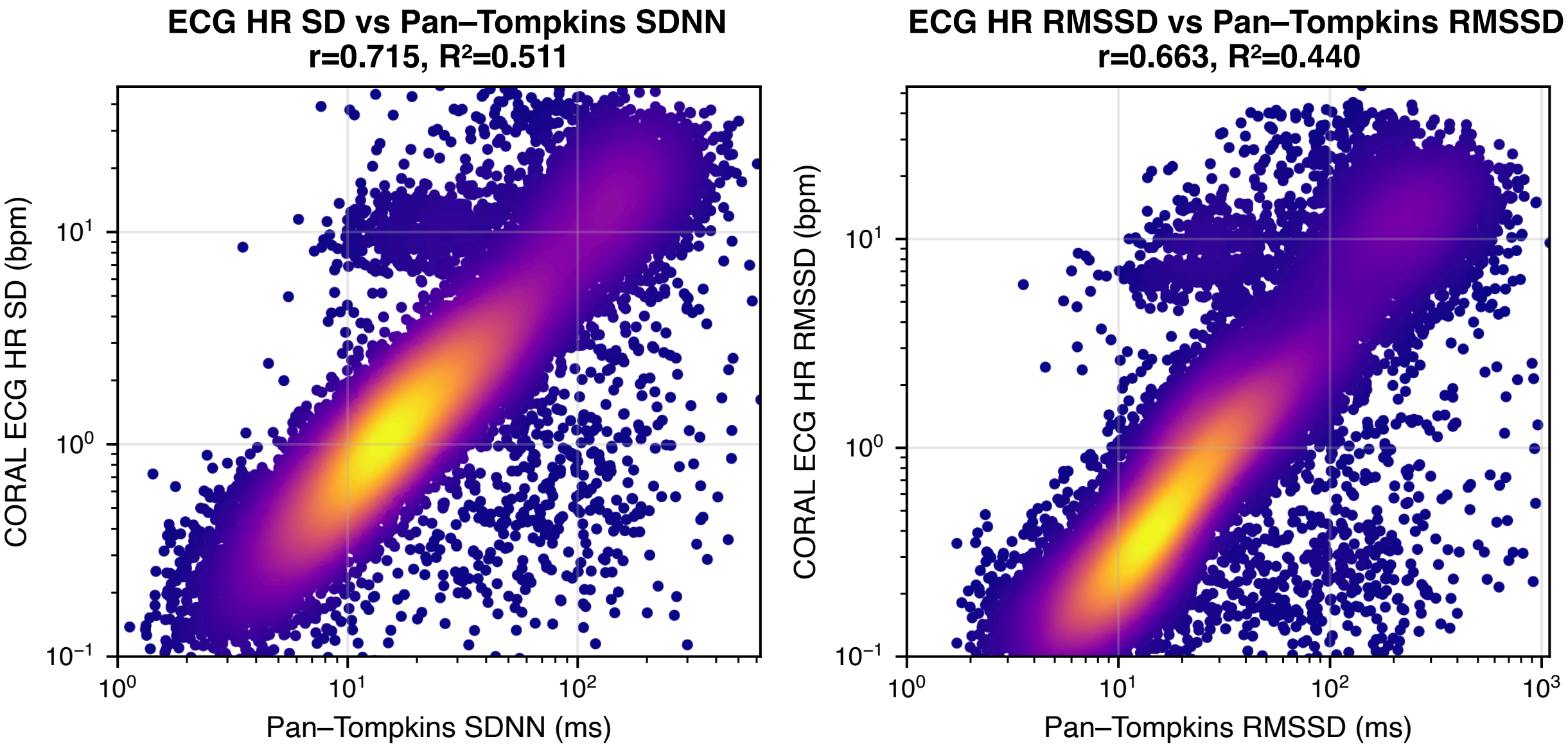

Figure S7: CORAL ECG HRV against Pan–Tompkins RR HRV (reference detector) from PTB-XL.

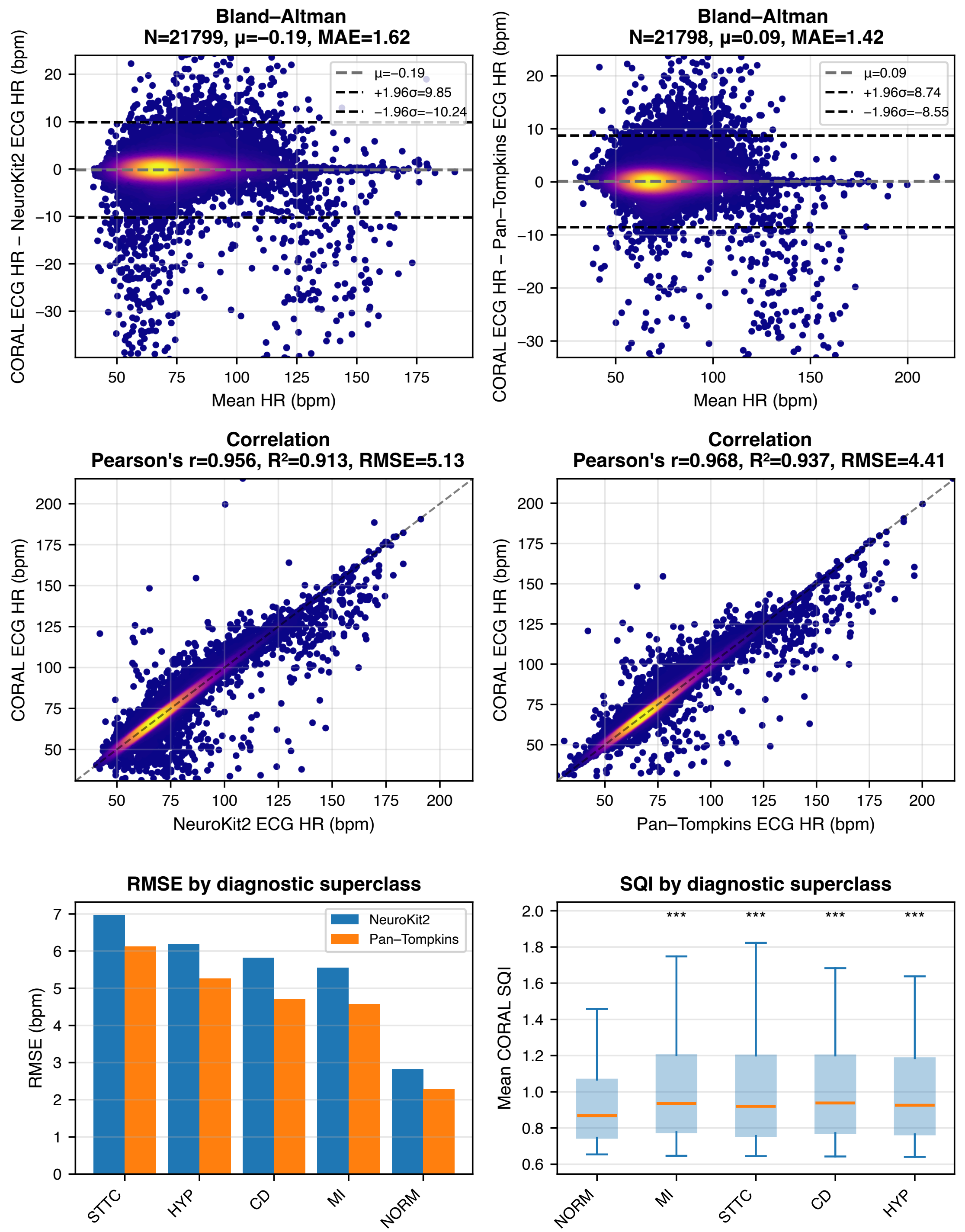

Figure S8: CORAL ECG HR against the NeuroKit2 and Pan–Tompkins reference detectors, with CORAL SQI by PTB-XL superclass. Single-superclass $n$ (NORM, MI, STTC, CD, HYP): 9069, 2532, 2400, 1708, 535. Kruskal–Wallis $H(4) = 195.754$, $p = 3.07 \times 10^{-41}$; ten pairwise two-sided Mann–Whitney U tests used Holm–Bonferroni FWER control. NORM was lower than MI, STTC, CD, and HYP (Cliff's $\delta = -0.143, -0.111, -0.128$, and $-0.116$, respectively; all Holm $p < 0.0001$); abnormal–abnormal $|\delta| \leq 0.027$ and Holm $p \geq 0.767$. Negative $\delta$ means lower NORM SQI; stars denote Holm $p$: * $< 0.05$, ** $< 0.01$, *** $< 0.001$. Abnormal-versus-NORM mean-SQI AUROC $= 0.577$ (95% record-bootstrap CI: 0.569–0.585).

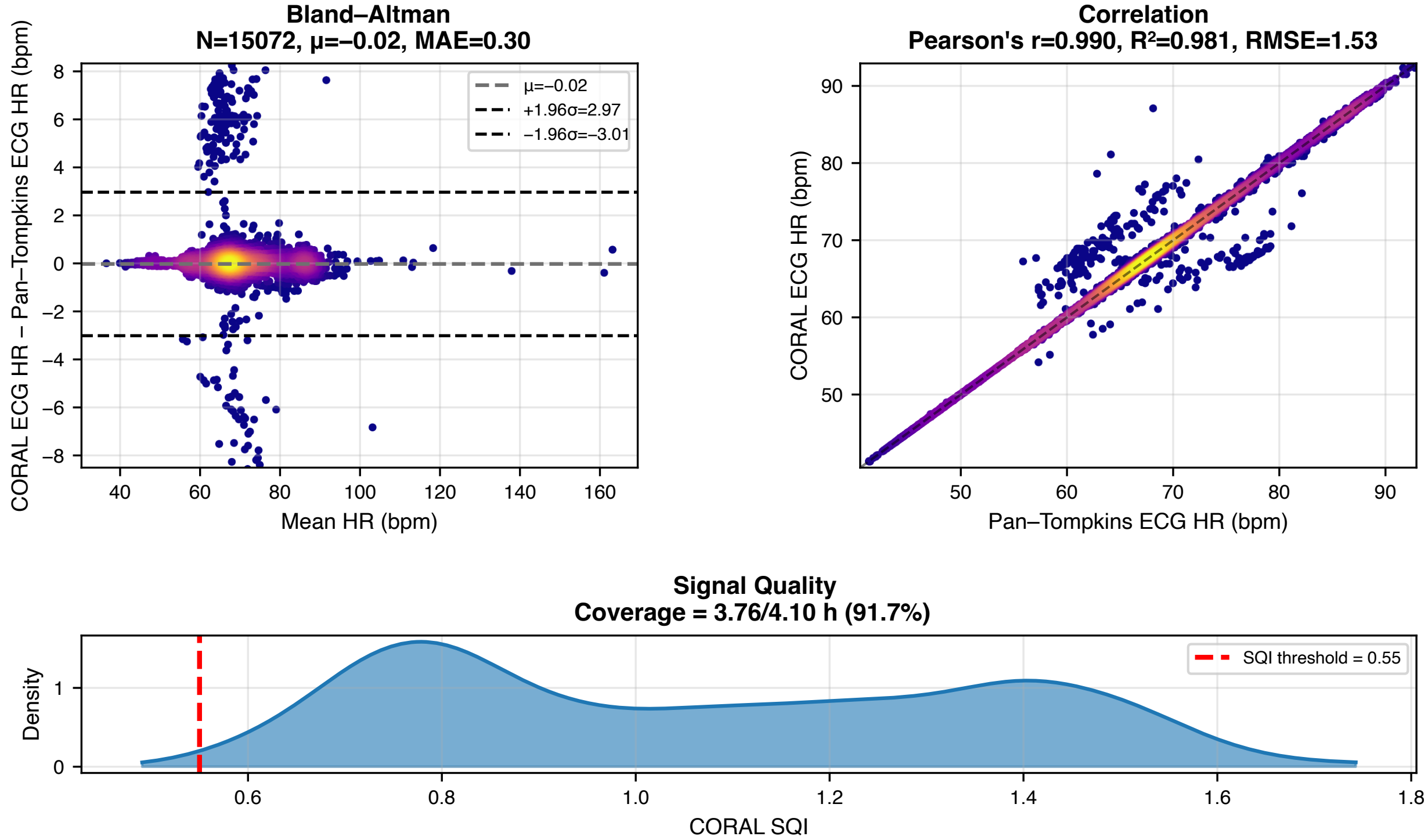

Figure S9: CORAL ECG HR performance against Pan–Tompkins ECG HR (reference detector) from ASXCG.

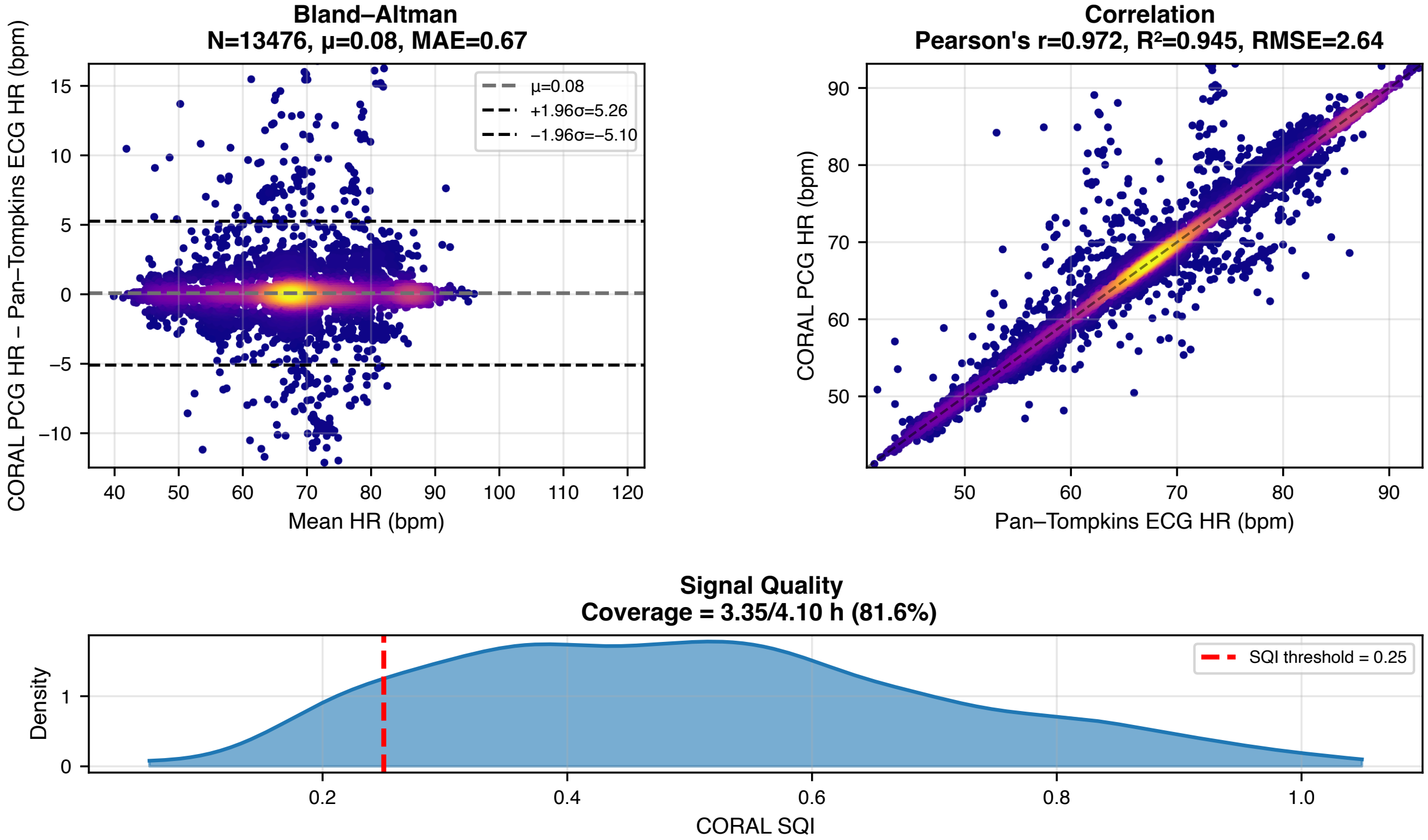

Figure S10: CORAL PCG HR performance against Pan–Tompkins ECG HR (separate reference signal) from ASXCG.

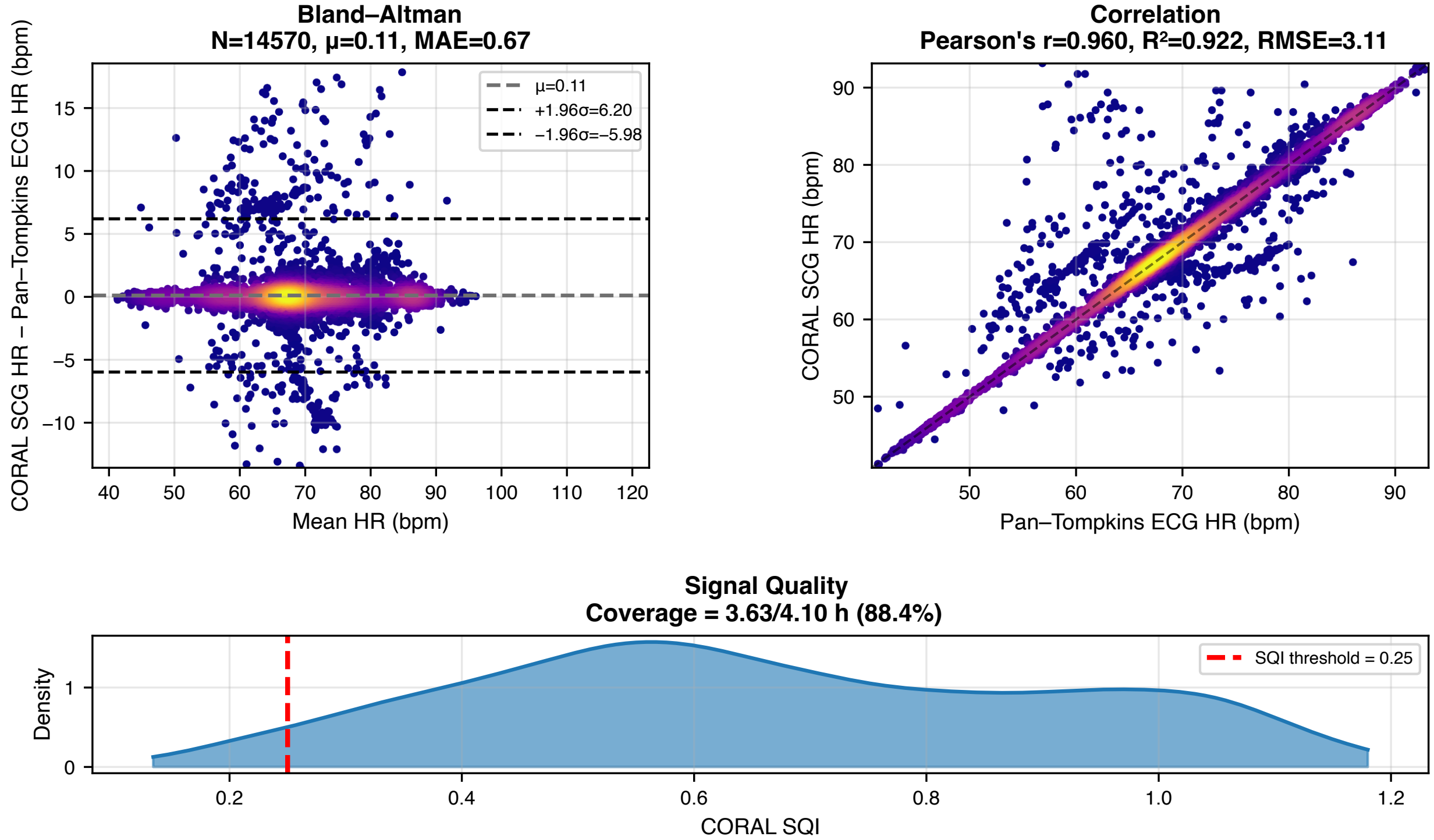

Figure S11: CORAL SCG HR performance against Pan–Tompkins ECG HR (separate reference signal) from ASXCG.

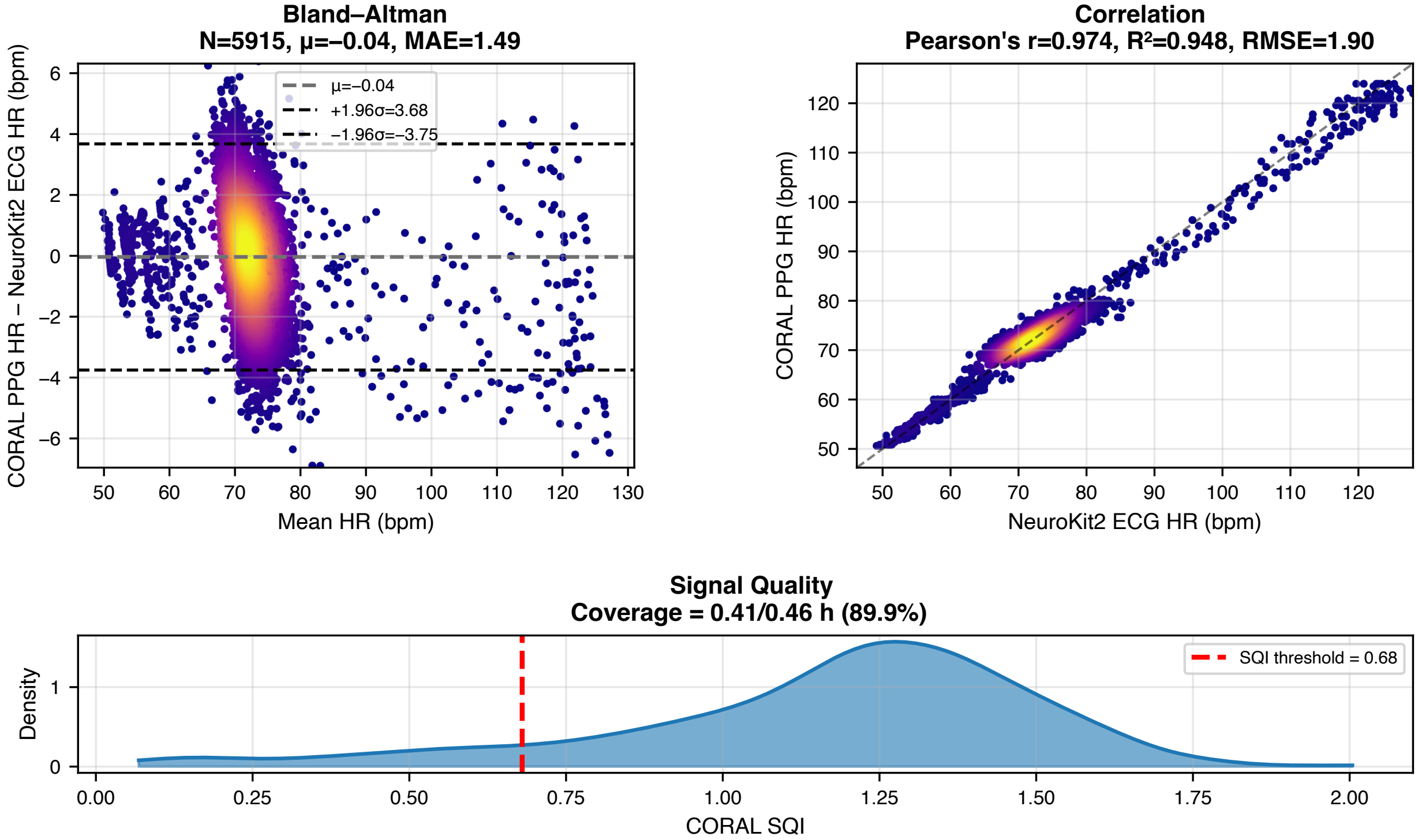

Figure S12: CORAL PPG HR performance against NeuroKit2 ECG HR (separate reference signal) from NIRS.

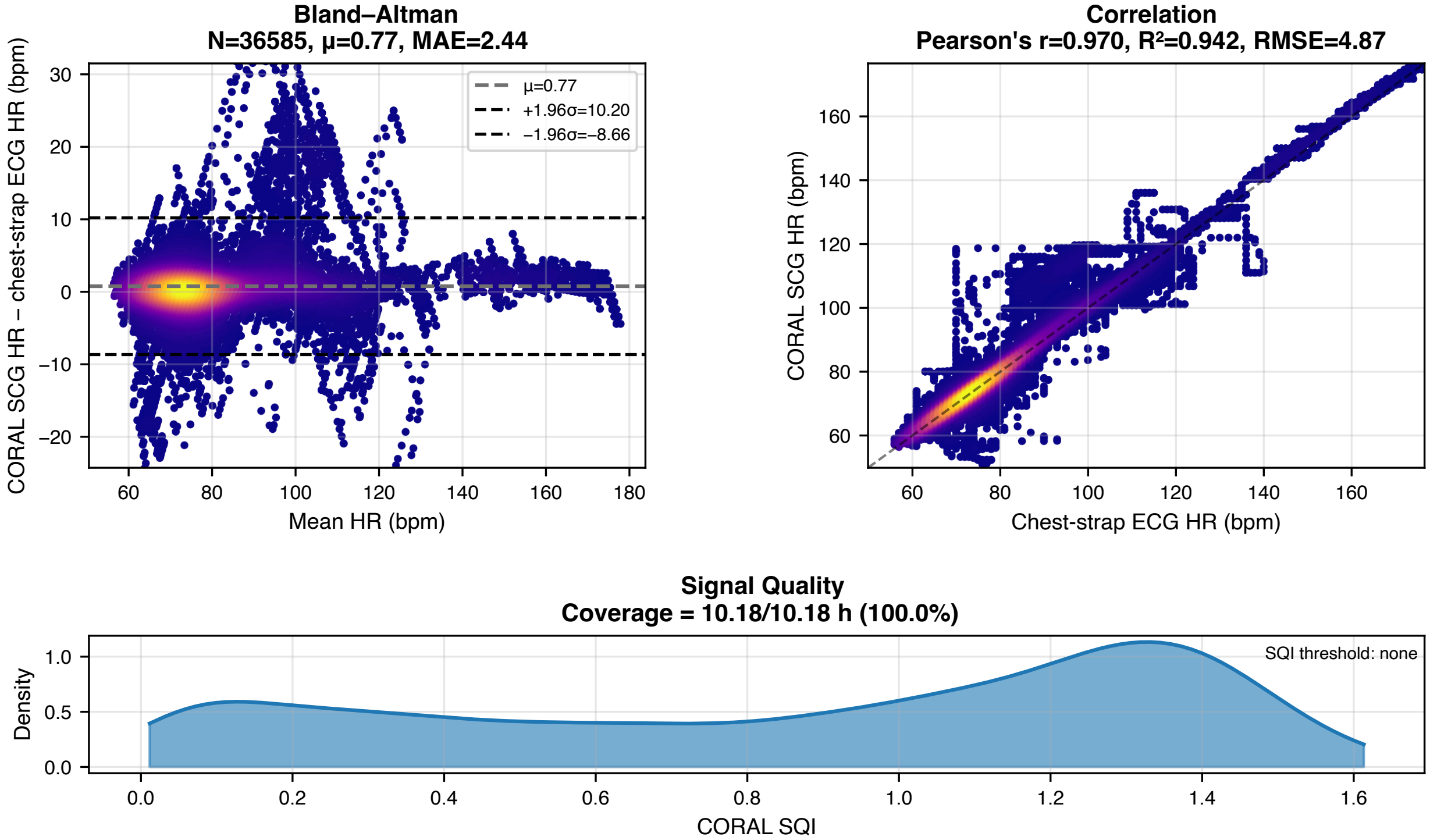

Figure S13: CORAL SCG HR performance against chest-strap ECG HR (separate reference signal) from AMB.

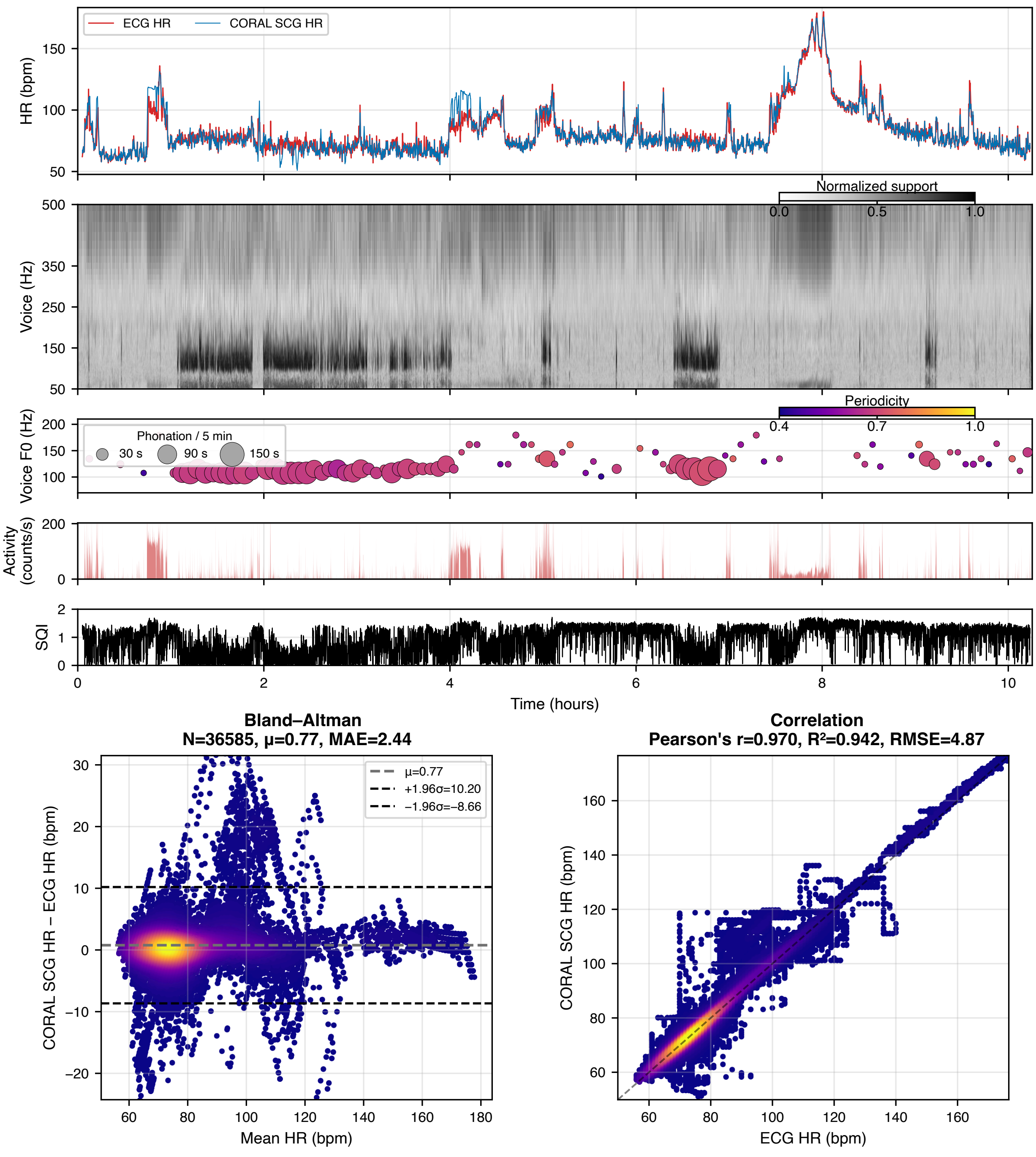

Figure S14: Continuous CORAL SCG HR and chest-strap ECG HR (separate reference signal), with a 50–500 Hz voice correloform, voice F0, activity, and SQI from AMB.

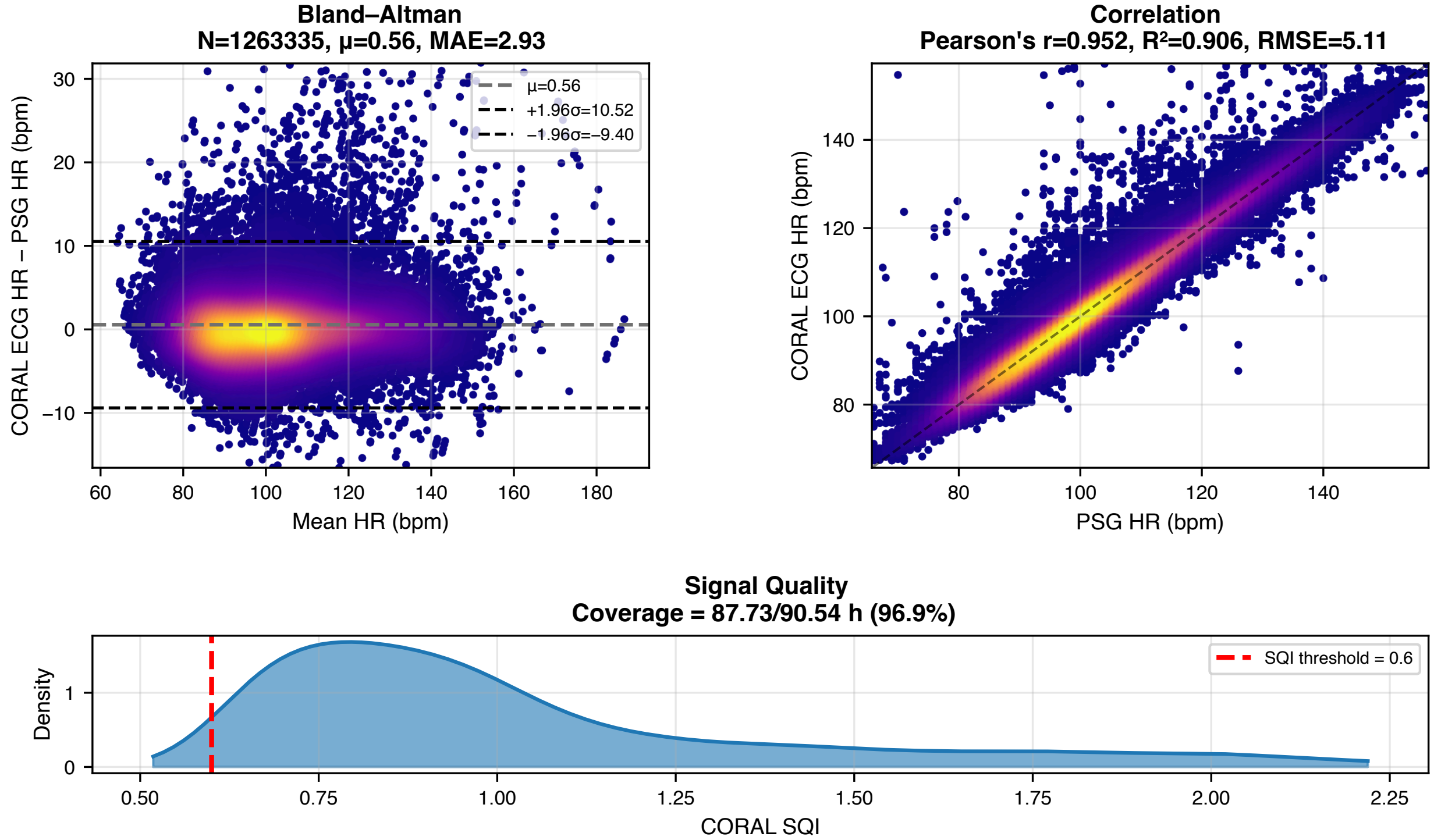

Figure S15: CORAL ECG HR performance against clinical PSG ECG HR (reference detector) from SLEEP.

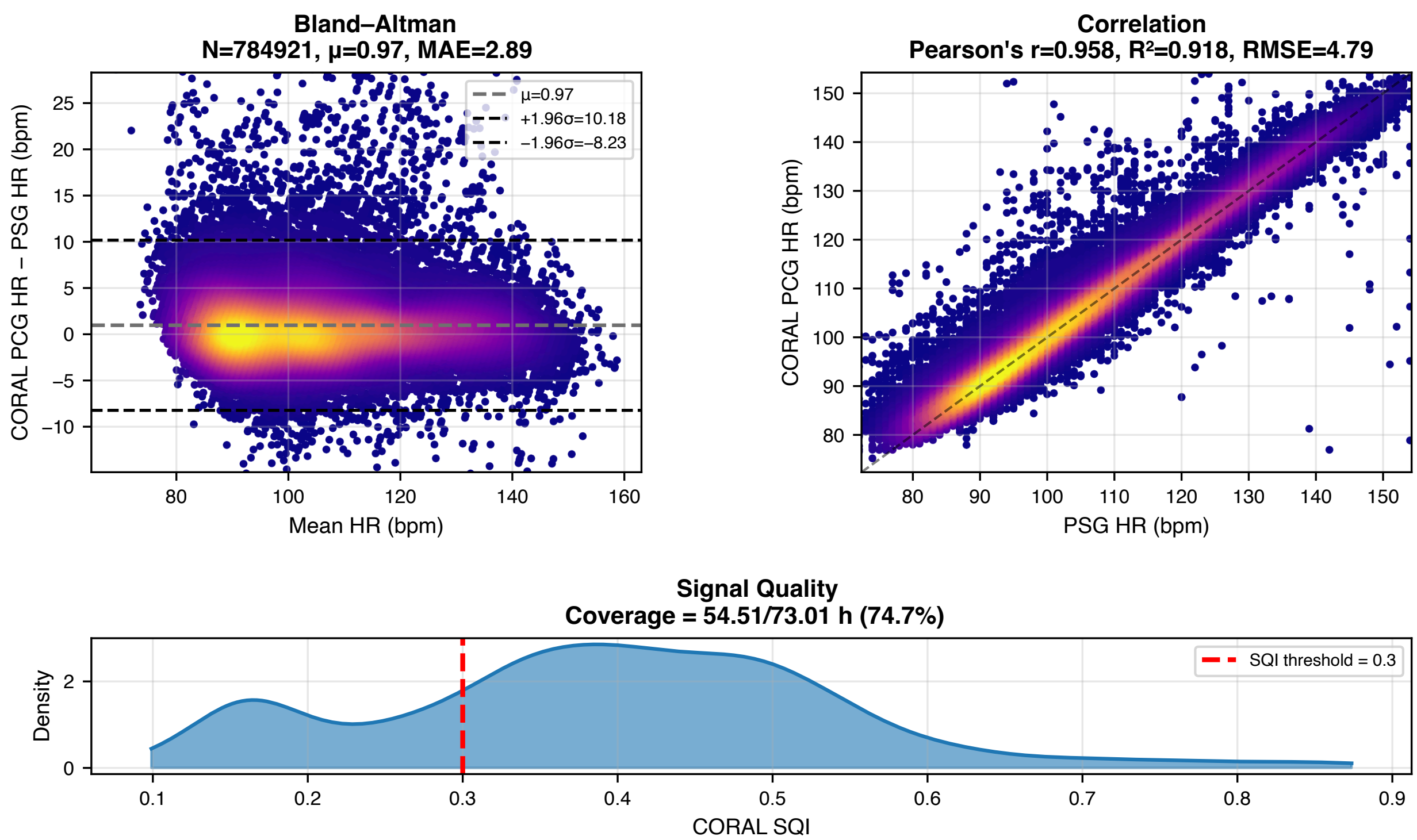

Figure S16: CORAL PCG HR performance against clinical PSG ECG HR (separate reference signal) from SLEEP.

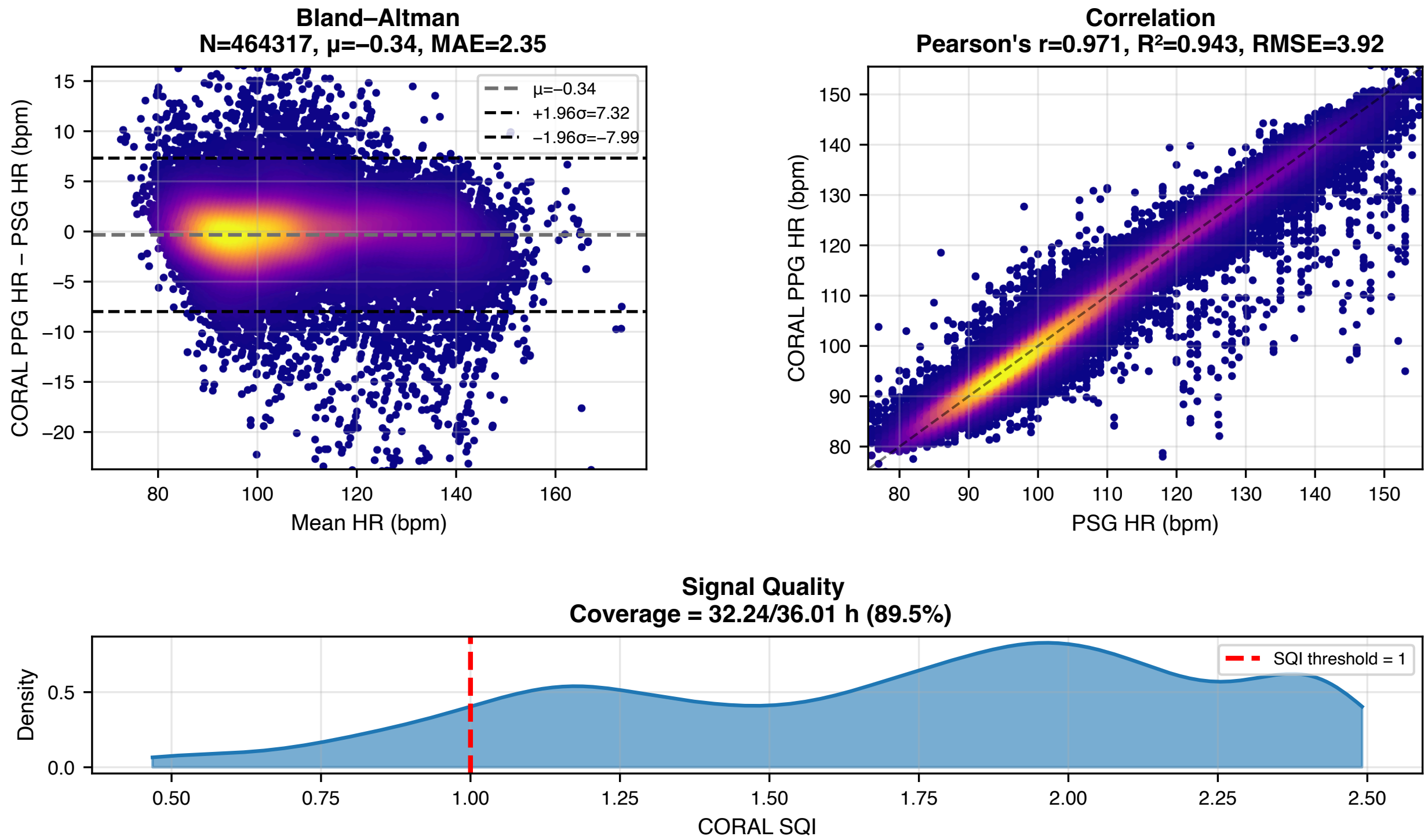

Figure S17: CORAL PPG HR performance against clinical PSG ECG HR (separate reference signal) from SLEEP.

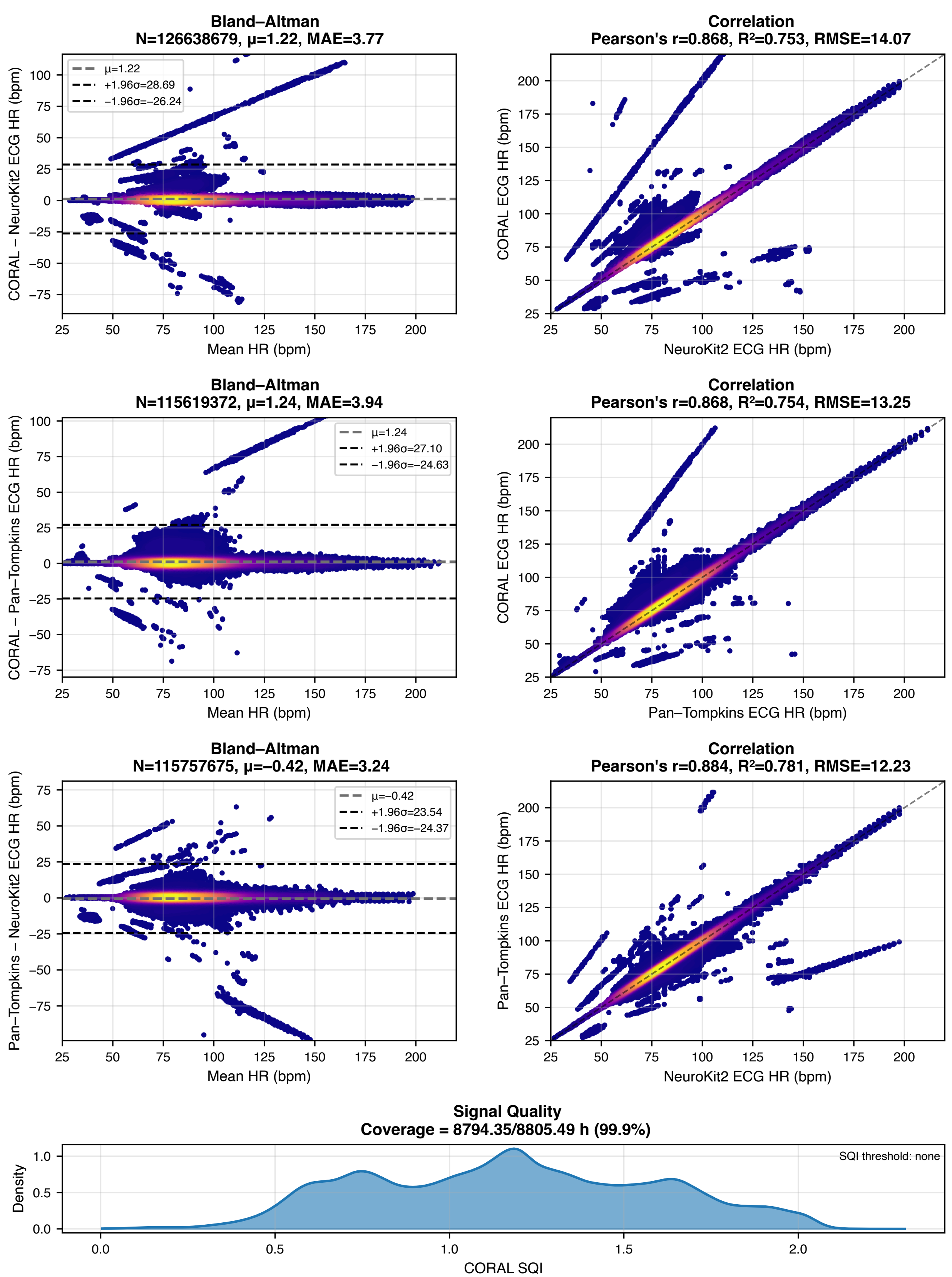

Figure S18: CORAL ECG HR against NeuroKit2 ECG HR and Pan–Tompkins ECG HR (reference detectors) from M4WDB. Half/double-rate disagreements associated with alternating RR intervals and bigeminy occur during 1.25% of paired time and contribute 25.2% of squared error.

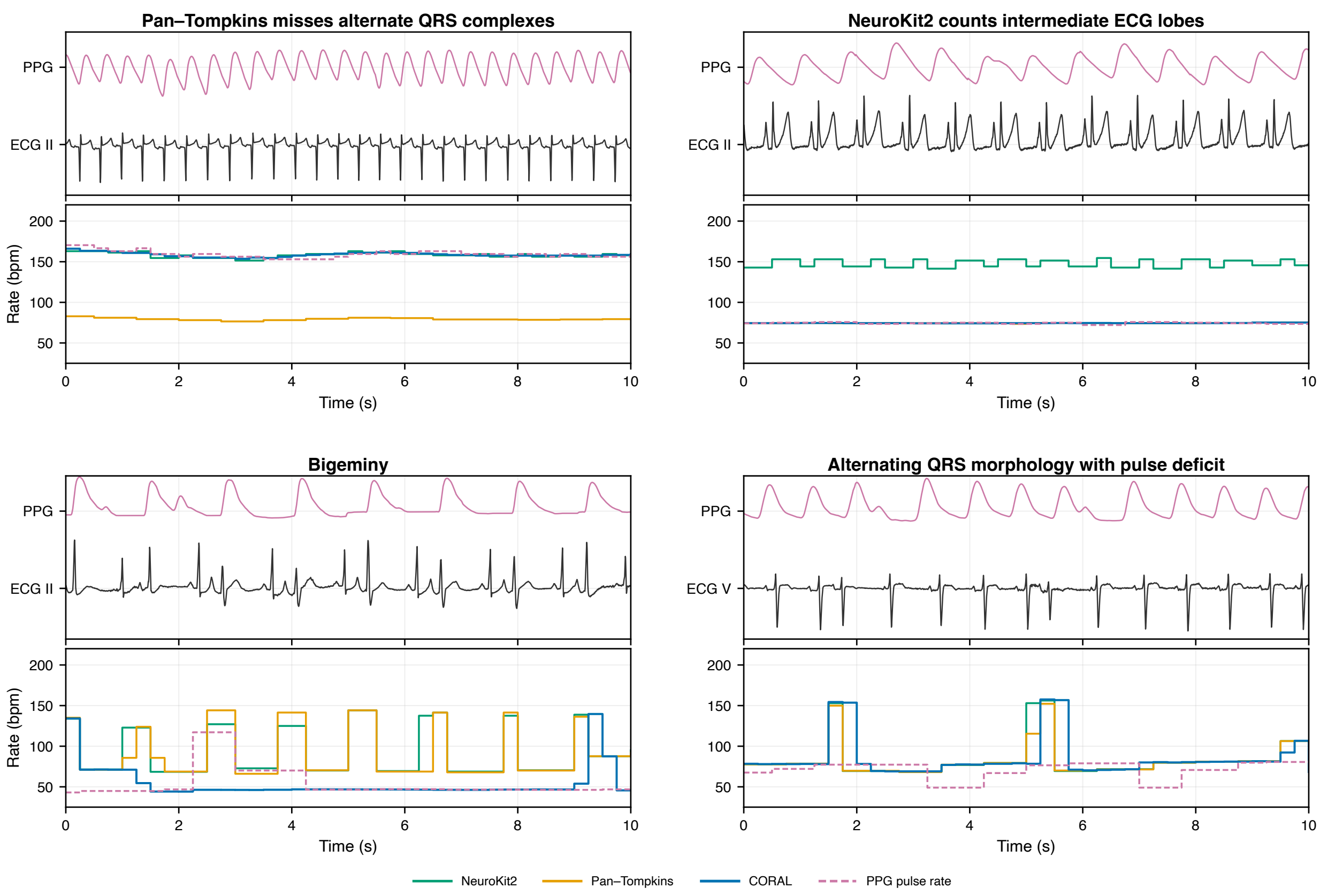

Figure S19: ECG detector disagreement and bigeminy examples with simultaneous PPG from M4WDB.

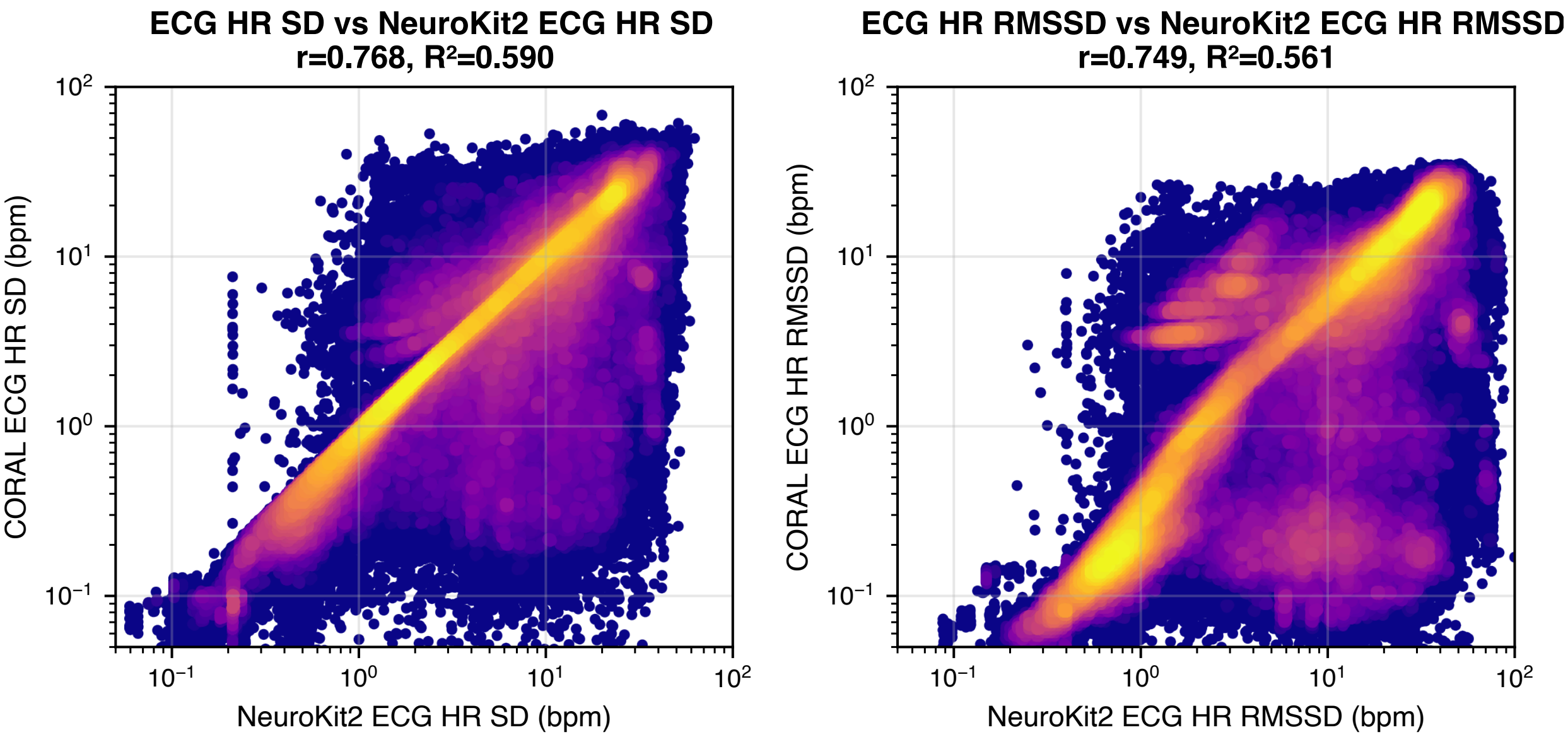

Figure S20: CORAL ECG HRV against NeuroKit2 ECG HRV over 60 s windows from M4WDB.

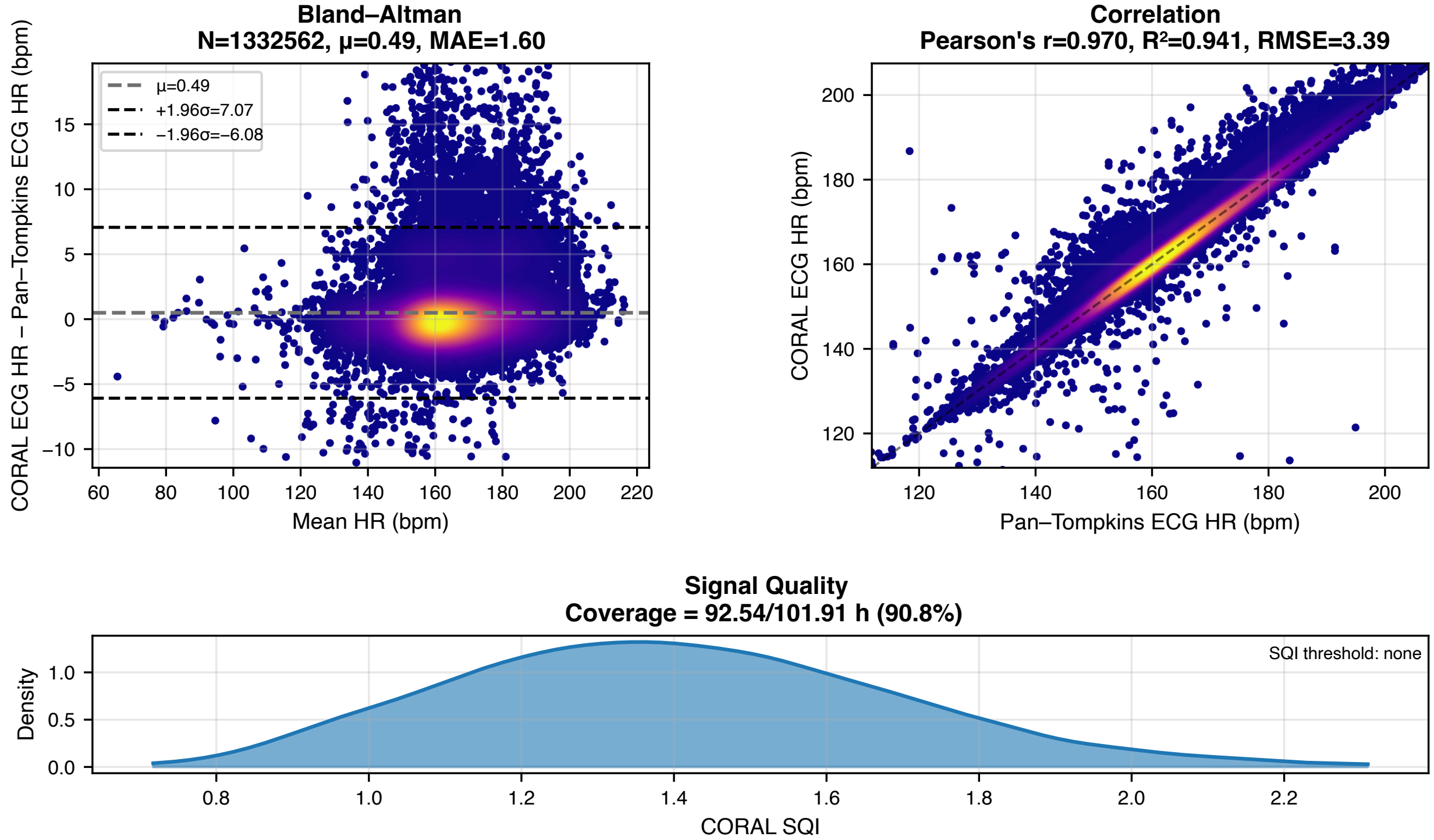

Figure S21: CORAL ECG HR performance against Pan–Tompkins ECG HR (reference detector) from ARC.

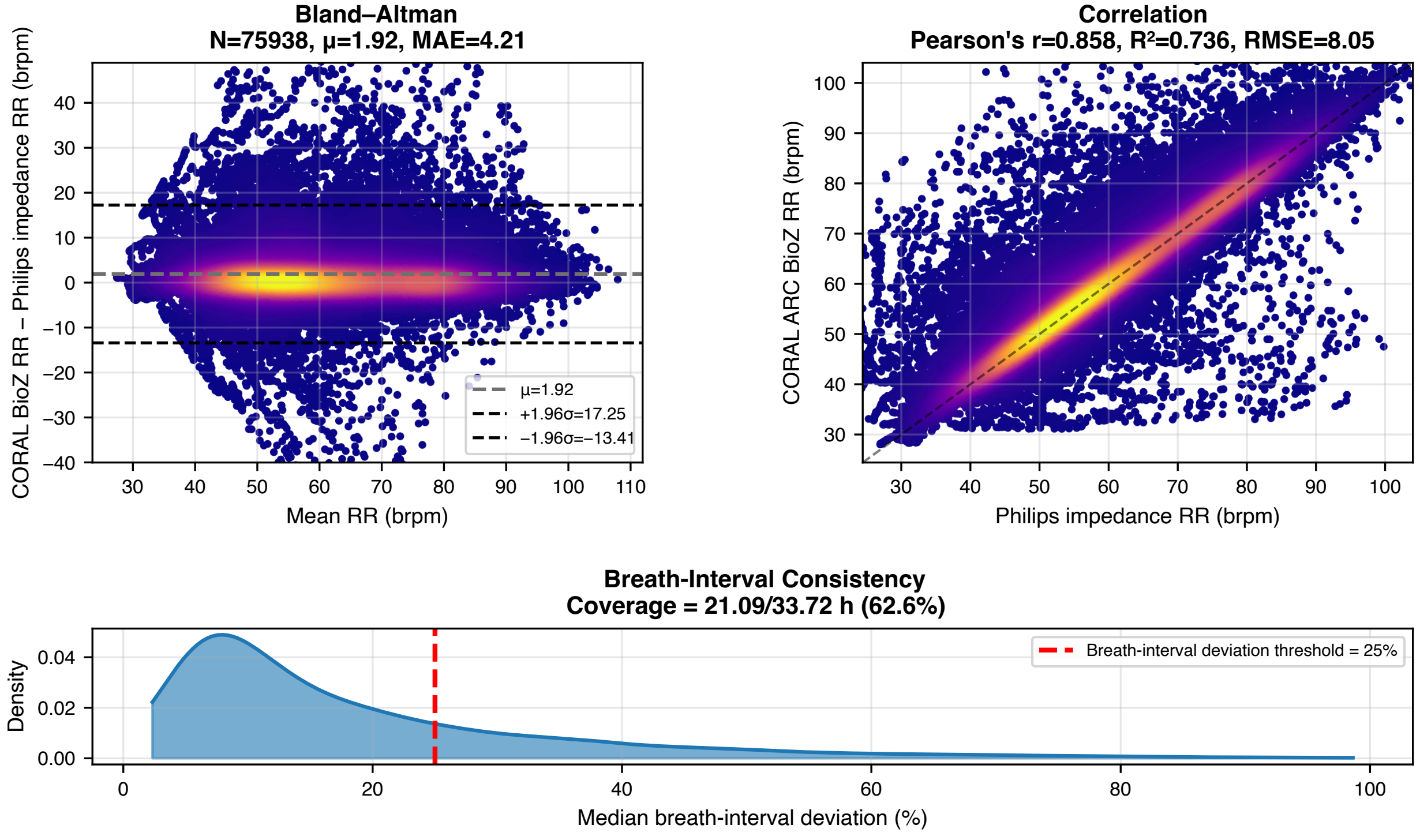

Figure S22: CORAL RR from ARC BioZ against RR from the simultaneously recorded wired Philips impedance waveform (separate reference signal).

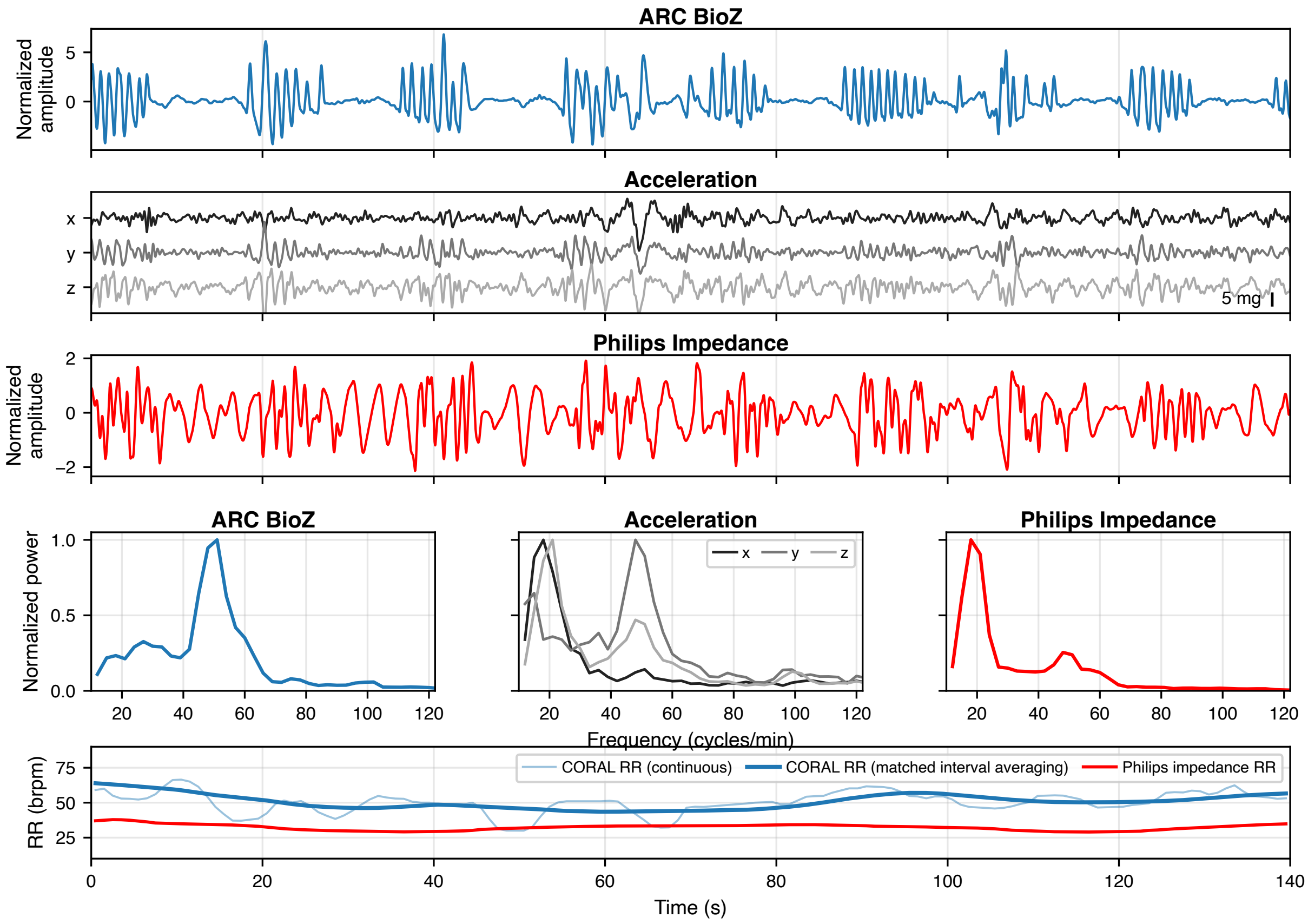

Figure S23: Periodic breathing in ARC BioZ and acceleration that is not dominant in the simultaneously recorded wired Philips impedance reference, with corresponding RR estimates.

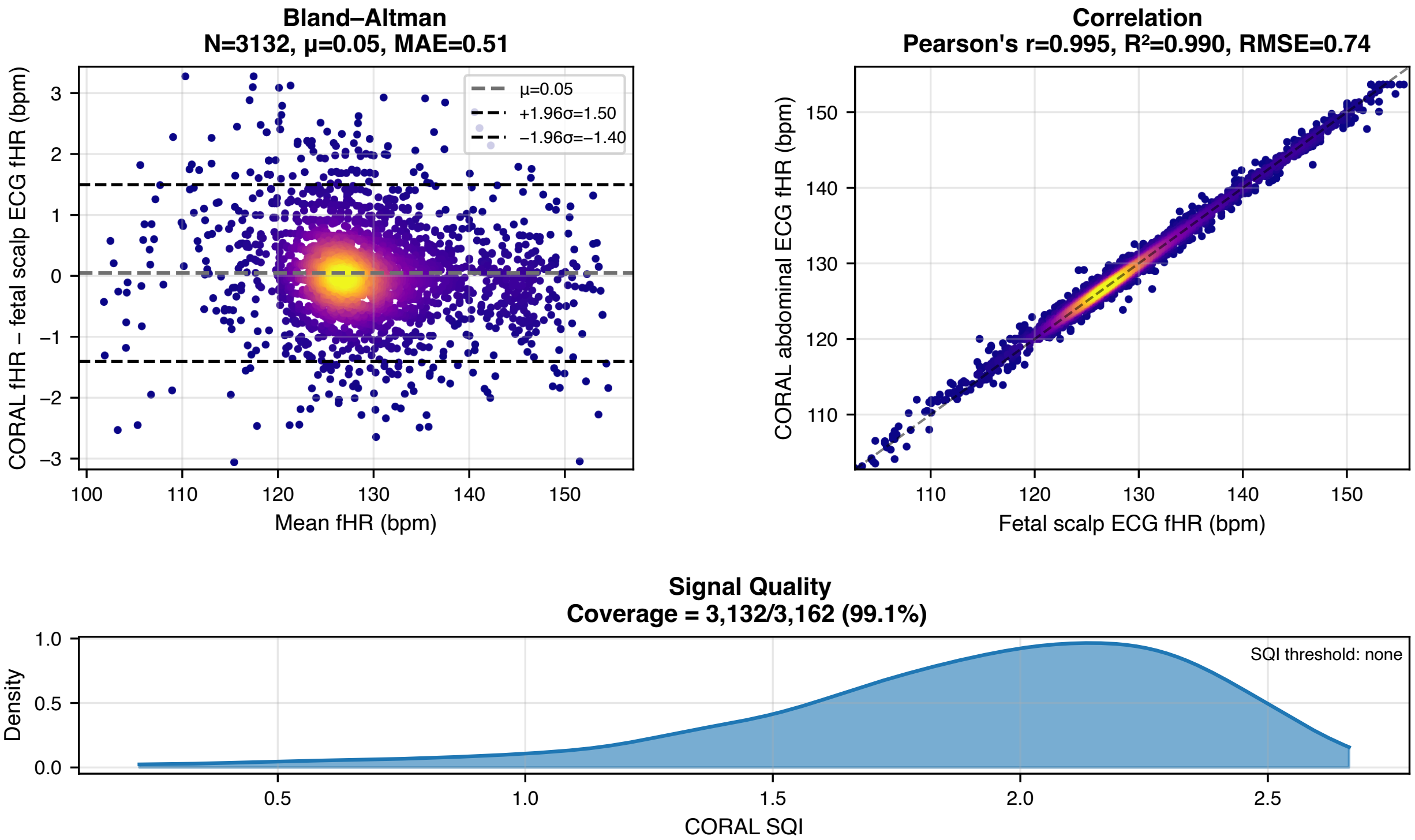

Figure S24: CORAL instantaneous fHR against fetal scalp ECG fHR (separate reference signal) from LABOR.

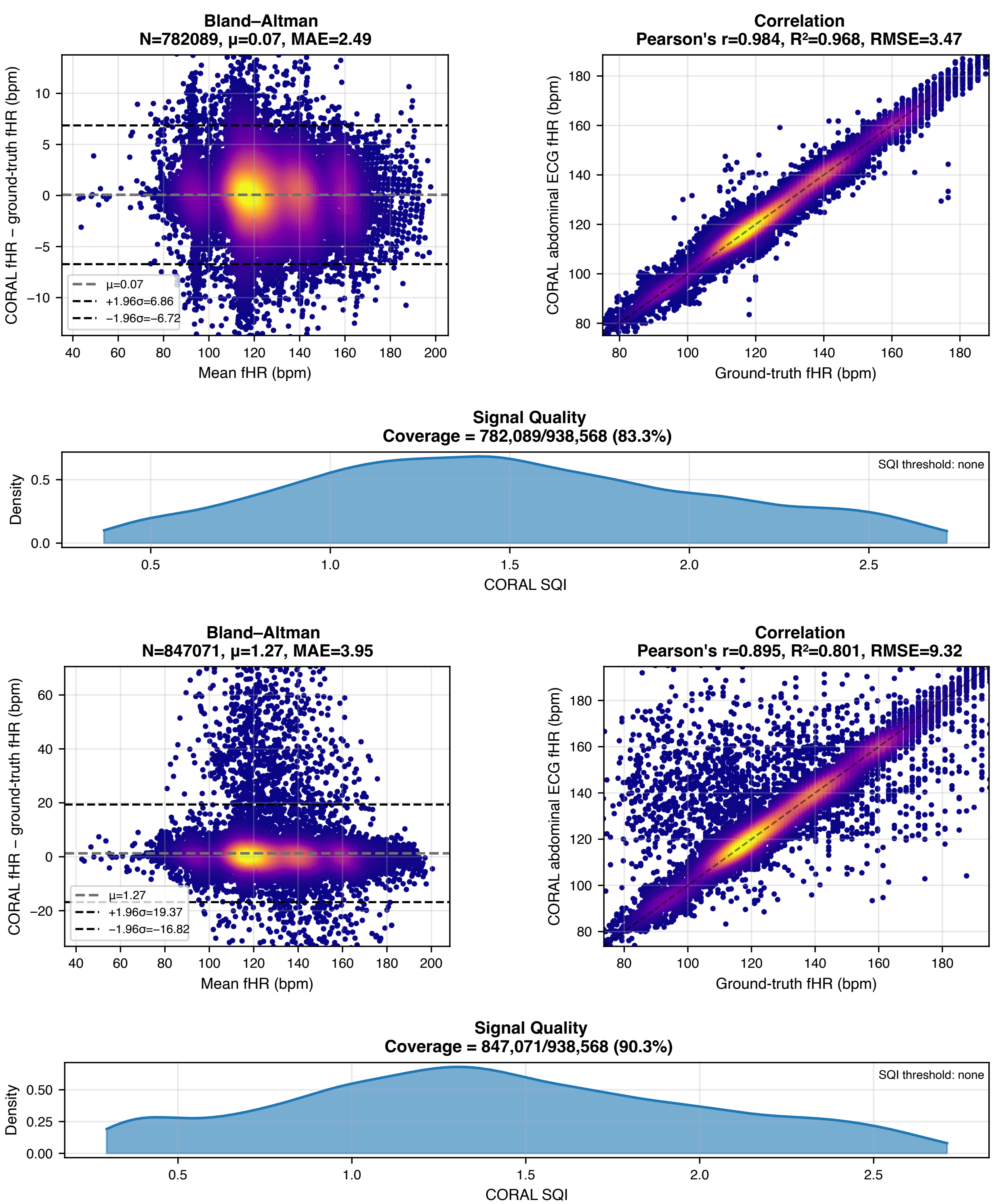

Figure S25: CORAL instantaneous fHR against synthetic ground-truth fHR from FSYNTH after automatic uterine-contamination exclusion (top) and with contractions retained (bottom).

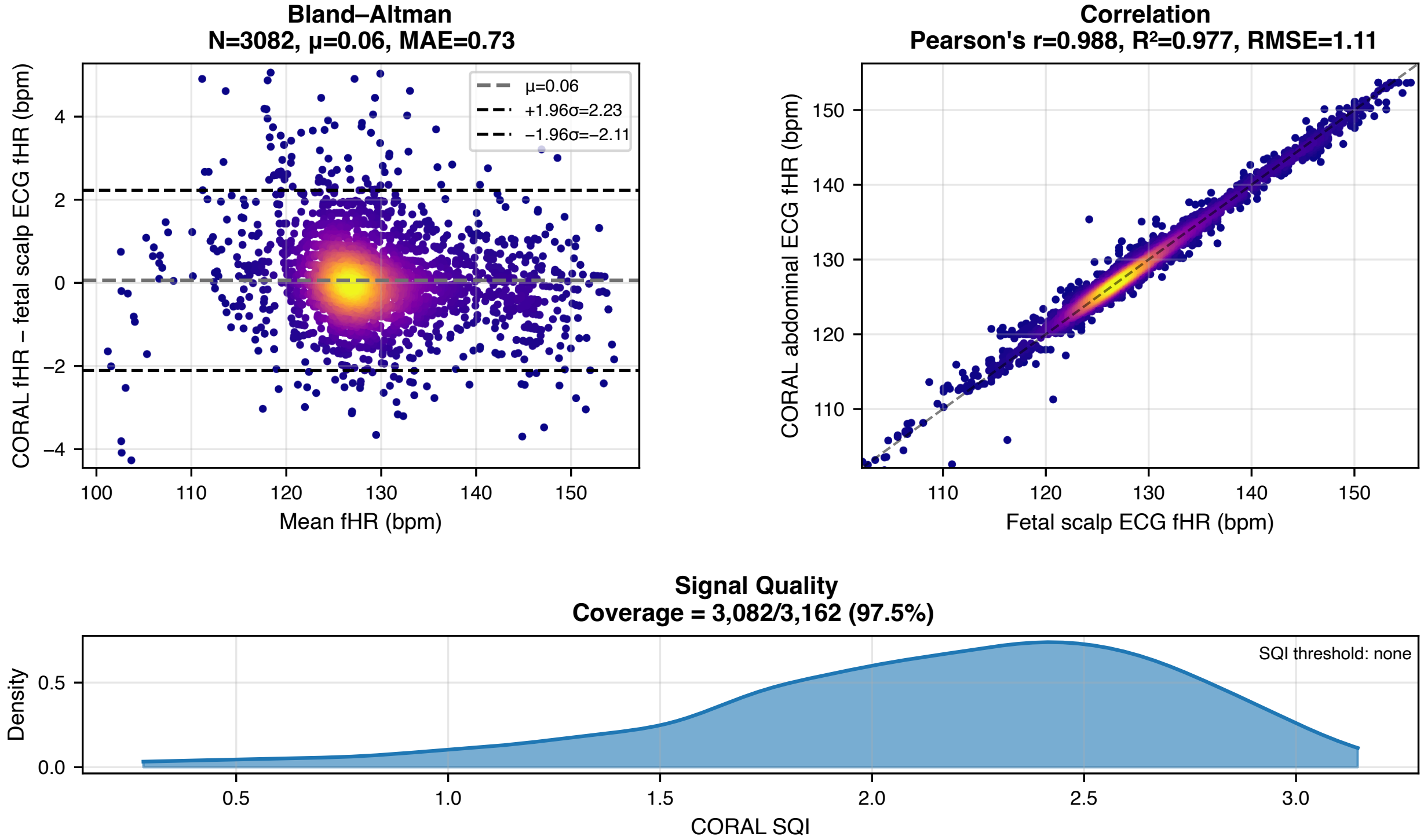

Figure S26: CORAL instantaneous fHR against fetal scalp ECG fHR (separate reference signal) from LABOR using FSYNTH parameters.

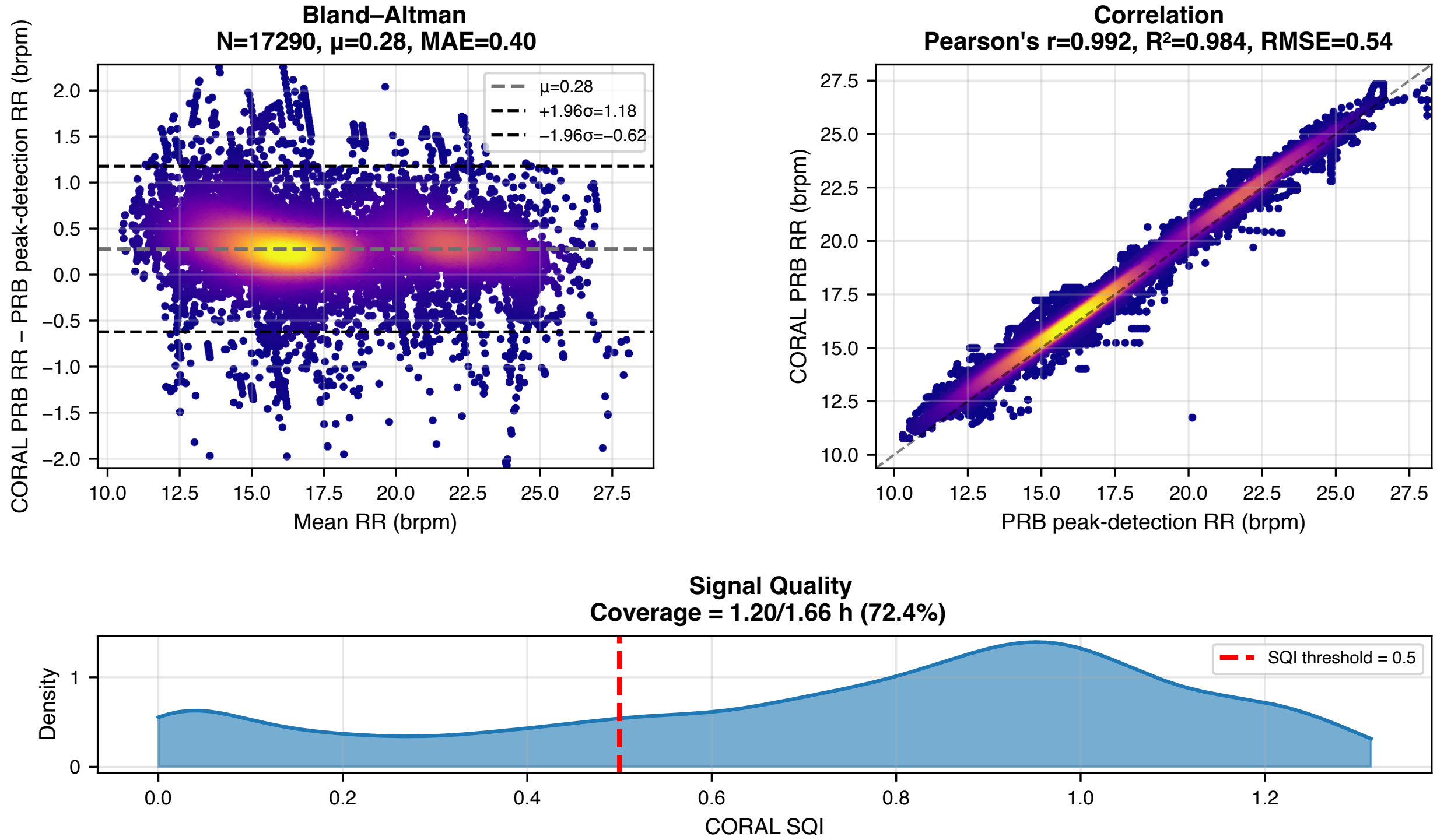

Figure S27: CORAL PRB RR performance against PRB peak-detection RR (reference detector) from CEBS.

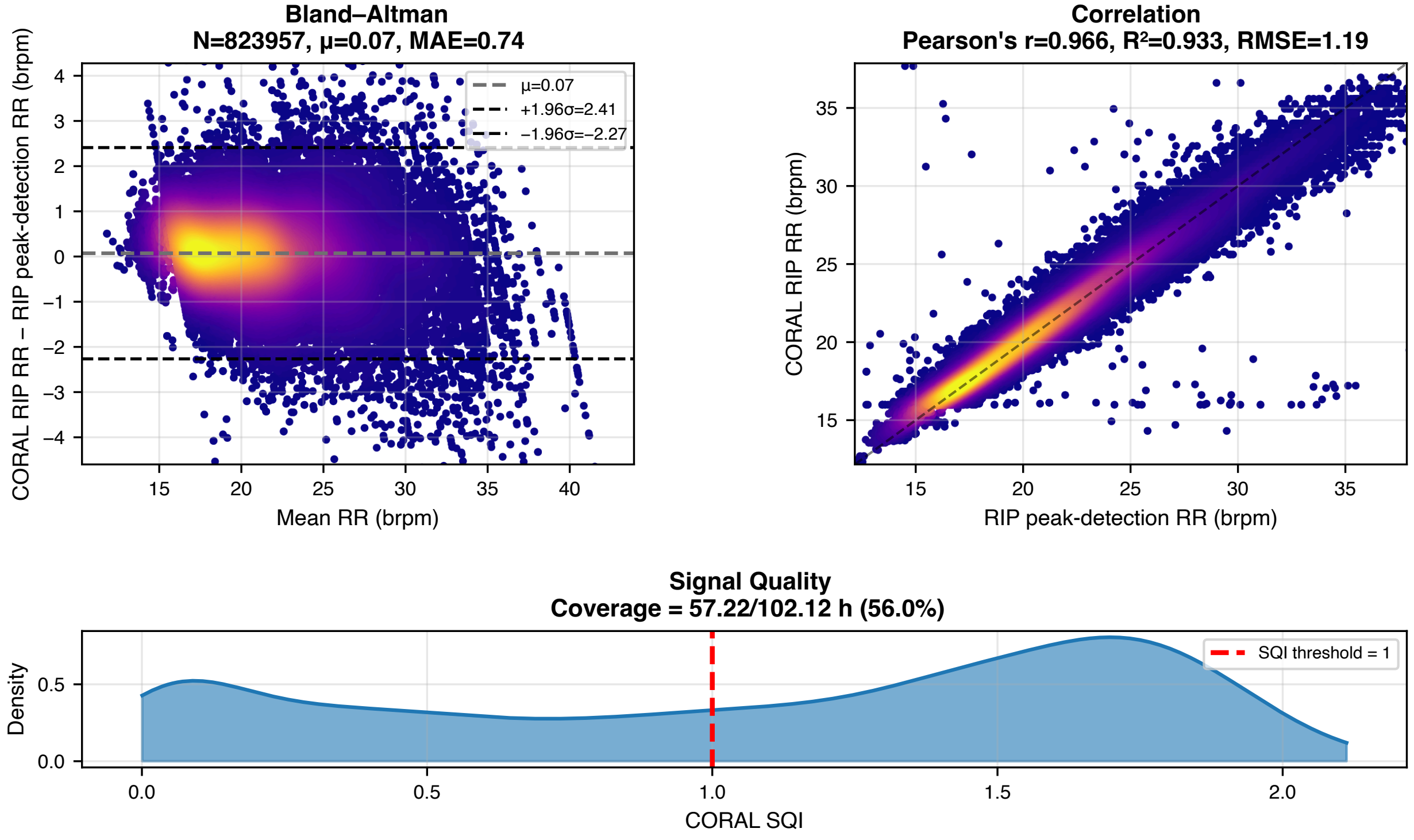

Figure S28: CORAL RIP RR performance against RIP peak-detection RR (reference detector) from SLEEP.

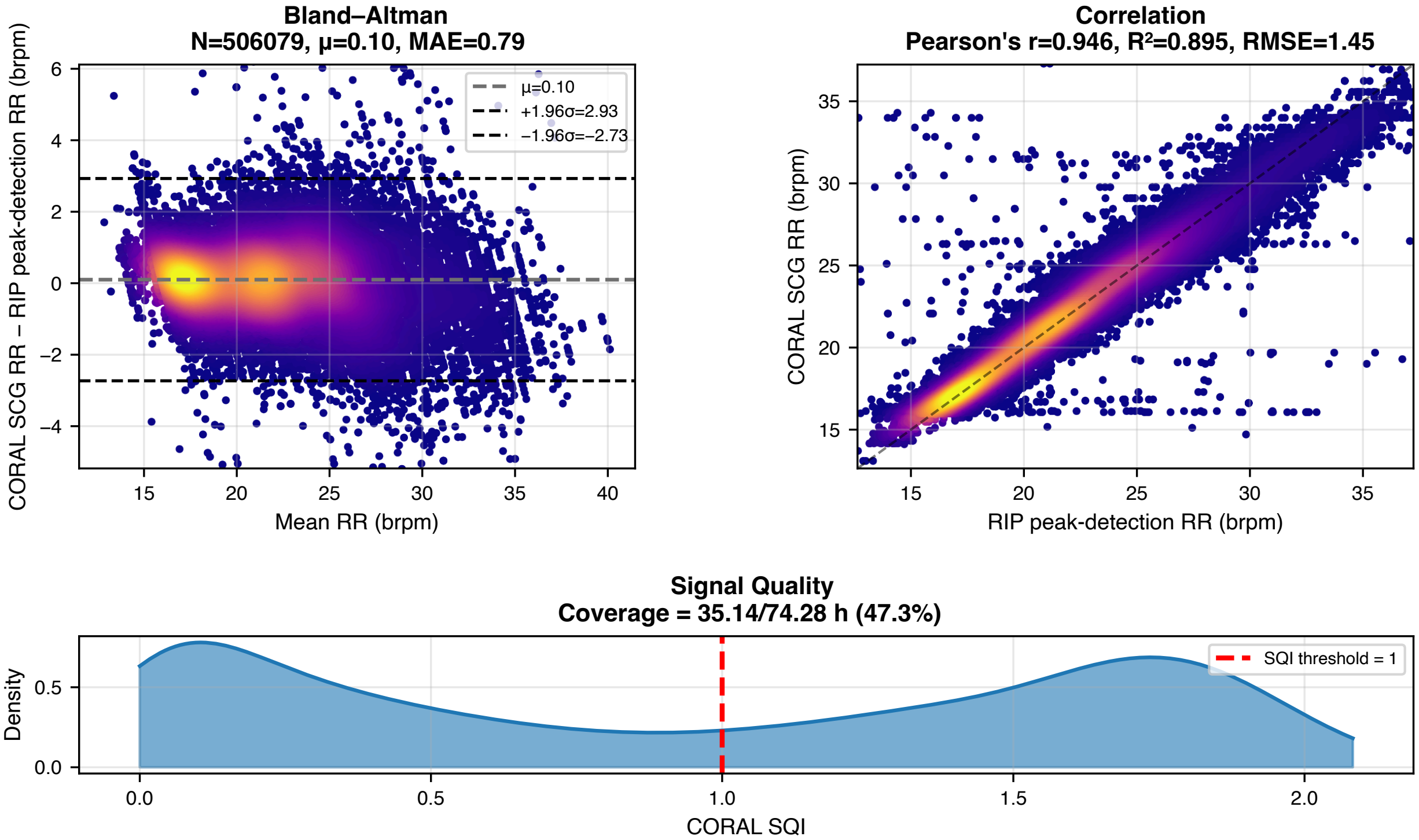

Figure S29: CORAL SCG RR performance against RIP peak-detection RR (separate reference signal) from SLEEP.

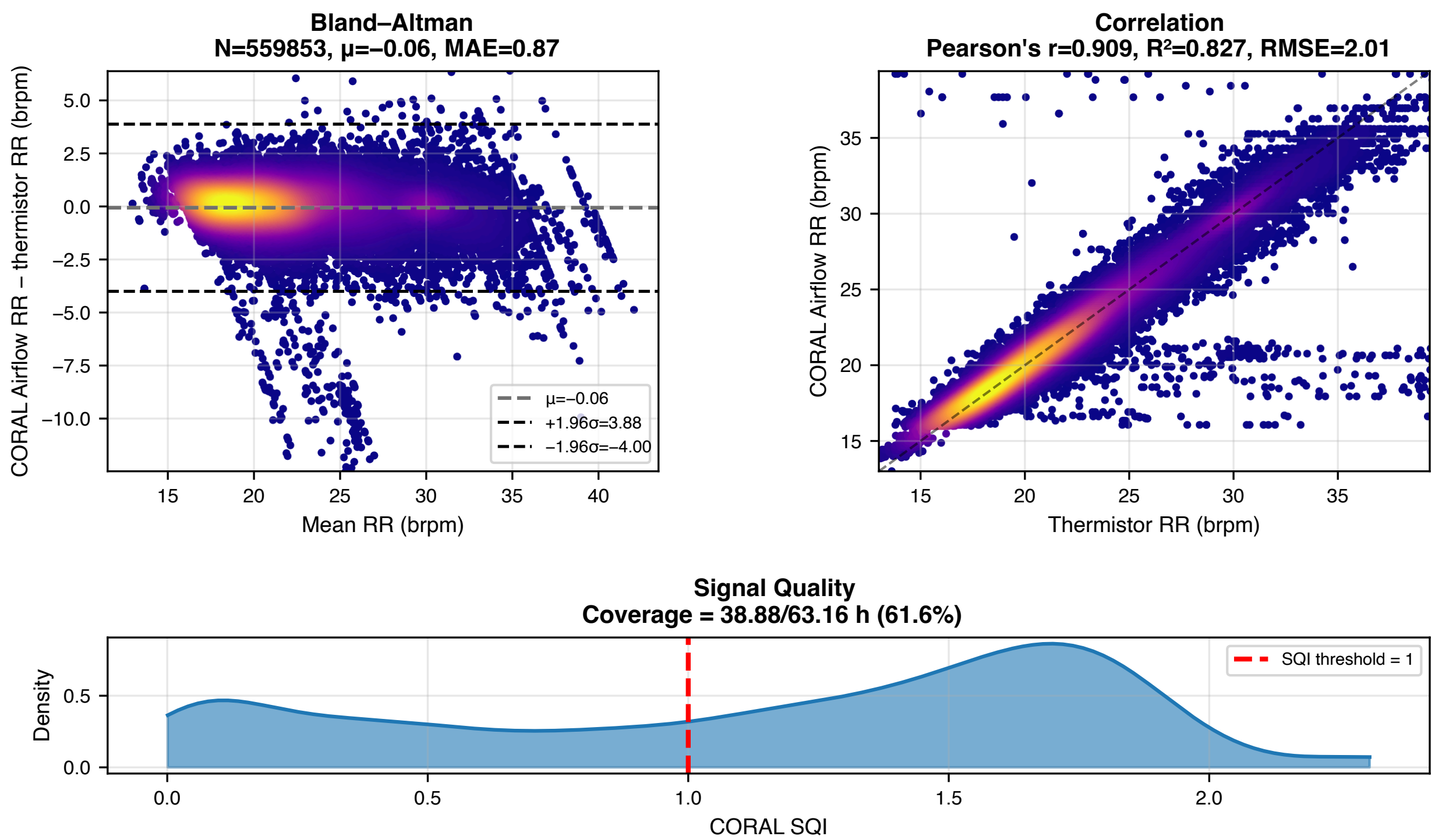

Figure S30: CORAL Airflow RR performance against thermistor breath-interval RR (reference detector) from SLEEP.

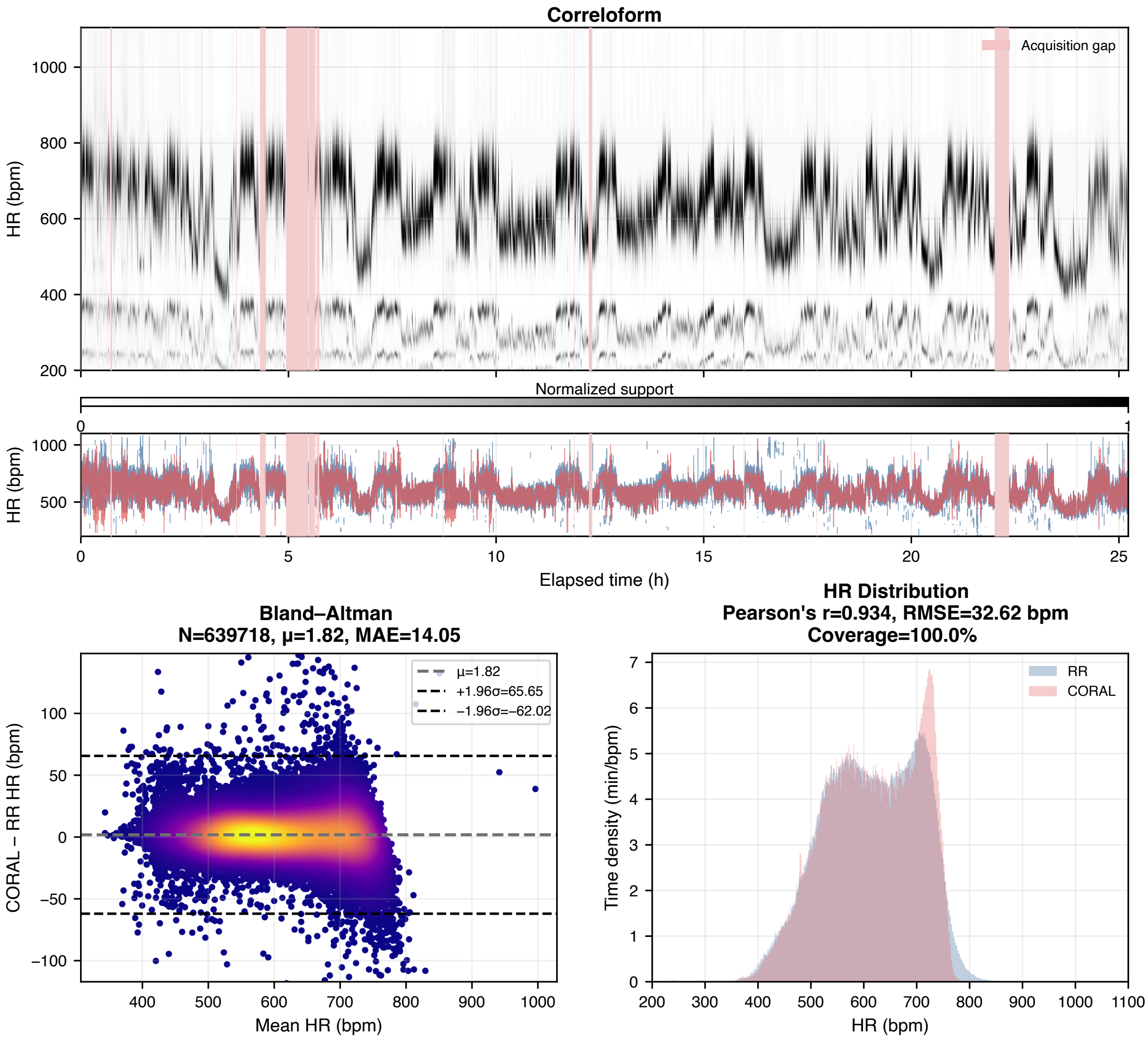

Figure S31: 24-hour ECG from a wireless mouse implant, with CORAL HR compared against mouse-adapted ECG envelope peak-detection HR (reference detector).